\documentclass[11pt,a4paper]{article}
\usepackage{jcappub,epsfig,multicol,bbm,latexsym,graphicx,subfigure}

\newcommand{\sv}{\ensuremath{\langle\sigma v\rangle}}

\begin{document}

\title{Systematic study of the uncertainties in fitting the 
cosmic positron data by AMS-02}

\author[a]{Qiang Yuan}
\author[a,1]{and Xiao-Jun Bi\note{For correspondence.}}
\affiliation[a]{Key Laboratory of Particle Astrophysics, Institute of High 
Energy Physics, Chinese Academy of Sciences, Beijing 100049, P.R. China}
\emailAdd{yuanq@ihep.ac.cn}
\emailAdd{bixj@ihep.ac.cn}

\abstract{
The operation of AMS-02 opens a new era for the study of cosmic ray 
physics with unprecedentedly precise data which are comparable with the 
laboratory measurements. The high precision data allow a quantitative 
study on the cosmic ray physics and give strict constraints on the nature 
of cosmic ray sources. However, the intrinsic errors from the theoretical 
models to interpret the data become dominant over the errors in the data. 
In the present work we try to give a systematic study on the uncertainties 
of the models to explain the AMS-02 positron fraction data, which shows 
the cosmic ray $e^+e^-$ excesses together with the PAMELA and Fermi-LAT 
measurements. The excesses can be attributed to contributions from the 
extra $e^+e^-$ sources, such as pulsars or the dark matter annihilation. 
The possible systematic uncertainties of the theoretical models considered 
include the cosmic ray propagation, the treatment of the low energy data,
the solar modulation, the $pp$ interaction models, the nuclei injection 
spectrum and so on. We find that in general a spectral hardening of the
primary electron injection spectrum above $\sim50-100$ GeV is favored by 
the data. Furthermore, the present model uncertainties may lead to a 
factor of $\sim2$ enlargement in the determination of the parameter 
regions of the extra source, such as the dark matter mass, annihilation 
rate and so on.
}

\keywords{dark matter theory, cosmic ray theory}

\arxivnumber{1408.2424}

\maketitle

\section{Introduction}

One of the biggest discoveries in the cosmic ray (CR) field in recent 
years is the excess of positrons found by PAMELA\footnote{Actually there 
were earlier hints on the cosmic positron excesses by HEAT 
\cite{1997ApJ...482L.191B} and AMS-01 \cite{2007PhLB..646..145A}, 
which did not attract enough attention due to the large errors.} 
\cite{2009Natur.458..607A,2010APh....34....1A}. The recent AMS-02 data 
confirmed PAMELA's discovery with very high precision and extended the 
energy range to $350$ GeV \cite{2013PhRvL.110n1102A}. 
The excess positrons require some kinds of ``primary'' positrons sources 
\cite{2009PhRvD..79b1302S,2013NuPhS.243...85M}, either the astrophysical
sources like nearby pulsars or exotic sources like dark matter (DM)
annihilation or decay. There had been many discussions about the origin 
of the positron excess in literature (see e.g., the reviews 
\cite{2009MPLA...24.2139H,2010IJMPD..19.2011F,2012APh....39....2S,
2012Prama..79.1021C,2013FrPhy...8..794B}).

Thanks to the high precision data from AMS-02, it is possible to perform 
very detailed study of the properties of the primary positron sources. 
Following our earlier works on the PAMELA results \cite{2010PhRvD..81b3516L,
2012PhRvD..85d3507L} we gave a quantitative study on the implications of 
the first AMS-02 positron fraction data in Ref. \cite{2015APh....60....1Y}. 
In that work we fitted the AMS-02 positron fraction as well as the 
PAMELA/Fermi-LAT/HESS electron (or total $e^+e^-$) spectra 
\cite{2011PhRvL.106t1101A,2009PhRvL.102r1101A,2010PhRvD..82i2004A,
2008PhRvL.101z1104A,2009A&A...508..561A} to determine the model parameters 
of both the extra sources and the CR background simultaneously. 
The fitting shows a tension to explain both the AMS-02 positron fraction 
and the Fermi total $e^+e^-$ spectrum simultaneously with the conventional 
background and extra source model, which works well in the PAMELA era 
\cite{2015APh....60....1Y}. Such a result was confirmed by several other 
studies \cite{2013JCAP...11..026J,2013PhRvD..88b3013C,2013PhRvD..87l3003M}
and was supported by the AMS-02 preliminary results of the total 
$e^+e^-$ spectrum \cite{2013ICRC-AMS02}. 

Although the AMS-02 data are precise enough, the theoretical framework to 
interpret the data still has large uncertainties, such as the uncertainties
from the CR propagation and the solar modulation. In the present work 
we will give a systematic study of such kinds of uncertainties. Especially 
we will show how the uncertainties will affect the determination of the 
model parameters of the extra sources to interpret the data.

One major uncertainty is the CR propagation. The propagation of CRs in the 
Galaxy is a diffusive process. The propagation parameters are determined 
by the secondary-to-primary ratio where the secondaries are generated 
through interactions between the CRs and the interstellar medium (ISM) 
during the propagation. However, the propagation parameters, usually 
determined by the Boron-to-Carbon (B/C) ratio, have relatively large 
uncertainties \cite{2010APh....34..274D,2011ApJ...729..106T}. The 
degeneracy between the propagation parameters is also strong due to the 
lack of high quality unstable-to-stable ratio of secondaries, such as 
Beryllium-10 to Beryllium-9 ($^{10}$Be/$^9$Be) ratio. Certainly, 
with the accumulation of the AMS-02 data the uncertainty of CR propagation 
is expected to be reduced significantly in future. 

Another major uncertainty comes from the complexity of modeling the 
low energy spectra of the $e^+e^-$. The reacceleration or convection during
the propagation, as well as the solar modulation, will affect the low
energy behavior of the particle spectrum. Sometimes people only select
the high energy data (e.g., $\gtrsim10$ GeV) in the studies to avoid
the complexity of the low energy spectra \cite{2014PhLB..728..250F,
2013JHEP...07..063I,2013ApJ...772...18L,2013JCAP...12..011G}. However, 
the results might be biased with limited range of the data. In this work 
we will show how the results may get affected with different selections 
of the low energy data. The solar modulation effect will be discussed too 
with proper approaches to address the uncertainties. Other effects, such 
as the inelastic hadronic interaction models or the injection proton 
spectrum, will affect the production of secondary $e^+e^-$. They will 
also be discussed in this work.

The properties of the extra sources themselves should also be a source of 
uncertainty. For example, the continuous or discrete distribution of the 
sources \cite{2009PhRvD..80f3005M}, and the burst-like or stable injection 
\cite{2010ApJ...710..958K} of the astrophysical sources will give different 
predictions of the resulting positron spectrum. For the DM scenario, the 
smooth DM distribution or the local DM clumps will also affect the model 
parameters to fit the data \cite{2009PhRvD..79j3513H,2009PhRvD..79l3517K}.
In the work we restrict our discussion with continuous source distribution 
of both the background and the extra source. For the DM annihilation scenario 
we consider the smooth distribution of the DM in the Milky Way halo and 
neglect the contribution from DM clumps \cite{2008A&A...479..427L,
2009PhRvD..80c5023B}. 

The paper is organized as follows. In Sec. \ref{ref} we describe the 
reference configuration as adopted in Ref. \cite{2013PhLB..727....1Y}. 
Sec. \ref{sys} presents the systematic uncertainties by considering 
different model configurations. We give some discussion about the results 
in Sec. \ref{dis} and finally give the conclusion in Sec. \ref{con}.

\section{Reference configuration}\label{ref}

The reference configuration is adopted following Ref. 
\cite{2013PhLB..727....1Y}. It is employed as a benchmark model to compare 
with different configurations and to illustrate the uncertainties. We 
briefly describe the major settings of the reference configuration here. 
The CR propagation is calculated with the public GALPROP 
package\footnote{http://galprop.stanford.edu/}\cite{1998ApJ...509..212S},
in which the secondary positron flux is predicted through the interaction
between the primary CR nuclei and the ISM. The propagation parameters are 
determined through a fit to the B/C ratio and $^{10}$Be/$^9$Be ratio in 
the diffusion reacceleration (DR) frame. The best fitting propagation 
parameters are $D_0|_{R_0=4\,{\rm GV}}=5.94\times10^{28}$ cm$^2$ s$^{-1}$, 
$\delta=0.377$, $v_A=36.4$ km s$^{-1}$ and $z_h=4.04$ kpc. 

The injection spectrum of the primary electrons are assumed to be a 
three-piece power-law function with respect to the rigidity,
\begin{eqnarray}
q(R)\propto\begin{cases}
(R/R_{\rm br,1}^{e})^{-\gamma_0},&R<R_{\rm br,1}^{e}\\
(R/R_{\rm br,1}^{e})^{-\gamma_1},&R_{\rm br,1}^{e}<R<R_{\rm br,2}^{e}\\
(R/R_{\rm br,2}^{e})^{-\gamma_2}(R_{\rm br,2}^{e}/R_{\rm br,1}^{e})^{-\gamma_1},&R>R_{\rm br,2}^{e}\ \  .
\end{cases}\label{eq:inj}
\end{eqnarray}
The first break $R_{\rm br,1}$ is at $\sim 4$ GV in order to fit the low 
energy data as well as the radio emission \cite{2011A&A...534A..54S}, 
and the second break $R_{\rm br,2}$ is at $O(100)$ GV to describe the 
spectral hardening. We consider two scenarios of the primary electron 
spectrum: {\it fittings I} for the cases without spectral hardening at 
$O(100)$ GV, and {\it fittings II} for the cases with a spectral hardening. 
The number of free parameters of {\it fittings II} will be larger by 2 (the 
break rigidity $R_{\rm br,2}^e$ and the spectral index above it, $\gamma_2$) 
than that of {\it fittings I}.

The second hardening of the background electron spectrum\footnote{This 
was originally motivated by the spectral hardening of CR nuclei around 
$\sim200$ GV by several experiments \cite{2007BRASP..71..494P,
2010ApJ...714L..89A,2011Sci...332...69A}. The most recent AMS-02 data 
about the proton and Helium spectra show, however, no remarkable structures 
below $\sim$TV \cite{2013ICRC-AMS02}. But the combination of AMS-02 data 
and CREAM data still shows a spectrum hardening at $\sim$TeV.}
was introduced to better fit the AMS-02 positron fraction and the Fermi 
total $e^+e^-$ spectrum \cite{2014PhLB..728..250F,2013PhRvD..88b3013C,
2013PhLB..727....1Y}. Other proposals, like charge asymmetry of the 
extra source \cite{2013PhRvD..87l3003M,2013JCAP...10..008F}, or the 
multi-component DM model \cite{2014PhRvD..89e5021G}, are also discussed to 
reconcile the tension. The requirement of more electrons at high energies 
may indicate the effect of discreteness of the primary CR sources 
\cite{2012MNRAS.421.1209T,2013A&A...555A..48B,2014JCAP...04..006D}.

The injection spectrum of the primary CR nuclei is adopted to be the
same form as shown in Eq. (\ref{eq:inj}). The parameters are derived 
through fitting to the PAMELA \cite{2011Sci...332...69A} and CREAM 
\cite{2010ApJ...714L..89A} proton spectra. The secondary electron and 
positron spectra can then be calculated in the same propagation model.
Since we are going to fit the $e^+e^-$ data, a free scale factor $c_{e^+}$ 
is multiplied to the predicted secondary $e^+e^-$ fluxes. Such a
factor is empirical and approximate. It may reflect the possible 
uncertainties of the overall fluxes of the secondary $e^+e^-$ fluxes, 
from e.g., the nuclear enhancement factor, ISM density distribution and 
so on \cite{2009NuPhB.813....1C}. Note that those uncertainties may not 
be exactly recovered by a constant factor \cite{2007APh....27..429H}. 
Since the discussion in this work will include the effects of spectral 
variations due to different propagation models and the hadronic interaction 
models, we do not employ further degree-of-freedom (dof, such as the form
$c_{e^+}E^p$ adopted in \cite{2009NuPhB.813....1C}) but keep such a simple 
re-scale factor, to avoid too much degeneracy of the secondary uncertainties.

Both the astrophysical pulsars and DM annihilation are considered as 
the extra sources contributing to the $e^+e^-$ excesses. The spatial 
distribution of the pulsars is adopted following the pulsar survey 
\cite{2004IAUS..218..105L}, and the injection spectrum of $e^+e^-$ from 
pulsars is parameterized as a single power-law with an exponential
cutoff, $E^{-\alpha}\exp(-E/E_c)$. The spectral index $\alpha$ is limited 
in the range of $0.6-2.2$ according to Fermi $\gamma$-ray observations 
\cite{2013ApJS..208...17A}. The case of multiple components of the 
pulsar contribution will not be discussed. Actually it has been shown that
adding more pulsars do not help to improve the fittings, although it may 
be physically realistic \cite{2013PhRvD..88b3001Y}. The single component
can effectively approach the sum of all the pulsars, although without the
fine structures \cite{2013PhRvD..88b3001Y}. As for the DM scenario, we 
adopt the annihilation channel $\tau^+\tau^-$ in this study. The hadronic 
channels will be constrained by the CR antiproton data
\cite{2010PhRvL.105l1101A,2009PhRvL.102g1301D}. Other leptonic channels, 
such as $\mu^+\mu^-$, $4e$, $4\mu$ and $4\tau$, may also work to fit the 
CR lepton data \cite{2010NuPhB.831..178M}. Given the purpose of this work
is to study to what extent the fitting results will be affect by various
kinds of systematics, the $\tau^+\tau^-$ channel is adopted as an 
illustration (see the Appendix for a comparison of the results for different 
annihilation channels). The density profile of DM is adopted to be 
Navarro-Frenk-White (NFW) distribution \cite{1997ApJ...490..493N}.
The spatial distribution of the $e^+e^-$ source does not sensitively 
affect the propagated results due to the limited propagation lengths of 
the high energy $e^+e^-$.

The solar modulation is treated by the center-force-field approximation 
\cite{1968ApJ...154.1011G}. A single modulation potential is assumed
for all the leptons in the reference configuration. Since the operation 
periods between PAMELA/Fermi-LAT and AMS-02 are moderately different in 
the solar cycle, we will test the case with two different modulation 
potentials for PAMELA/Fermi-LAT and AMS-02 data, respectively. In some 
works the charge-sign dependent modulation has been discussed to better 
describe the data \cite{1996ApJ...464..507C,2012AdSpR..49.1587D,
2013PhRvL.110h1101M}. We will also test the charge-dependent modulation 
effect by employing two different modulation potentials for $e^+$ and 
$e^-$ \cite{2009NJPh...11j5021B}, respectively.

\begin{figure*}[!htb]
\includegraphics[width=0.5\textwidth]{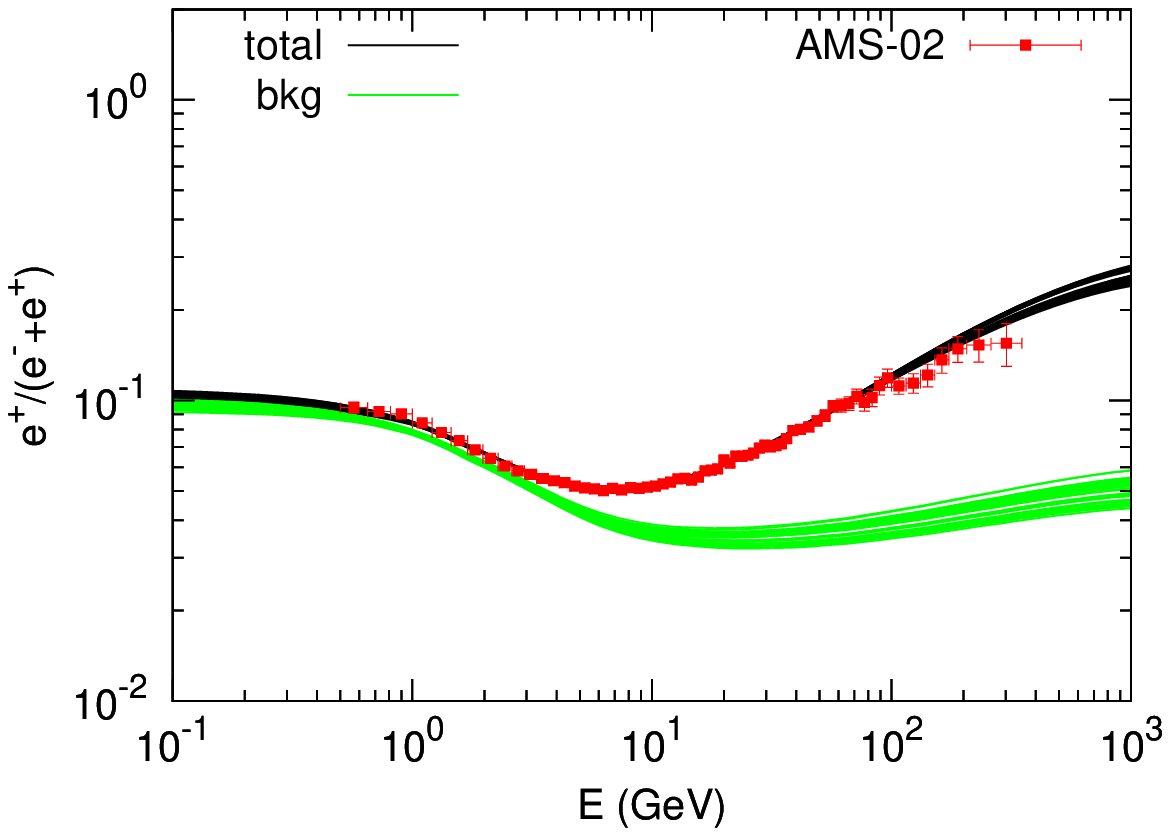}
\includegraphics[width=0.5\textwidth]{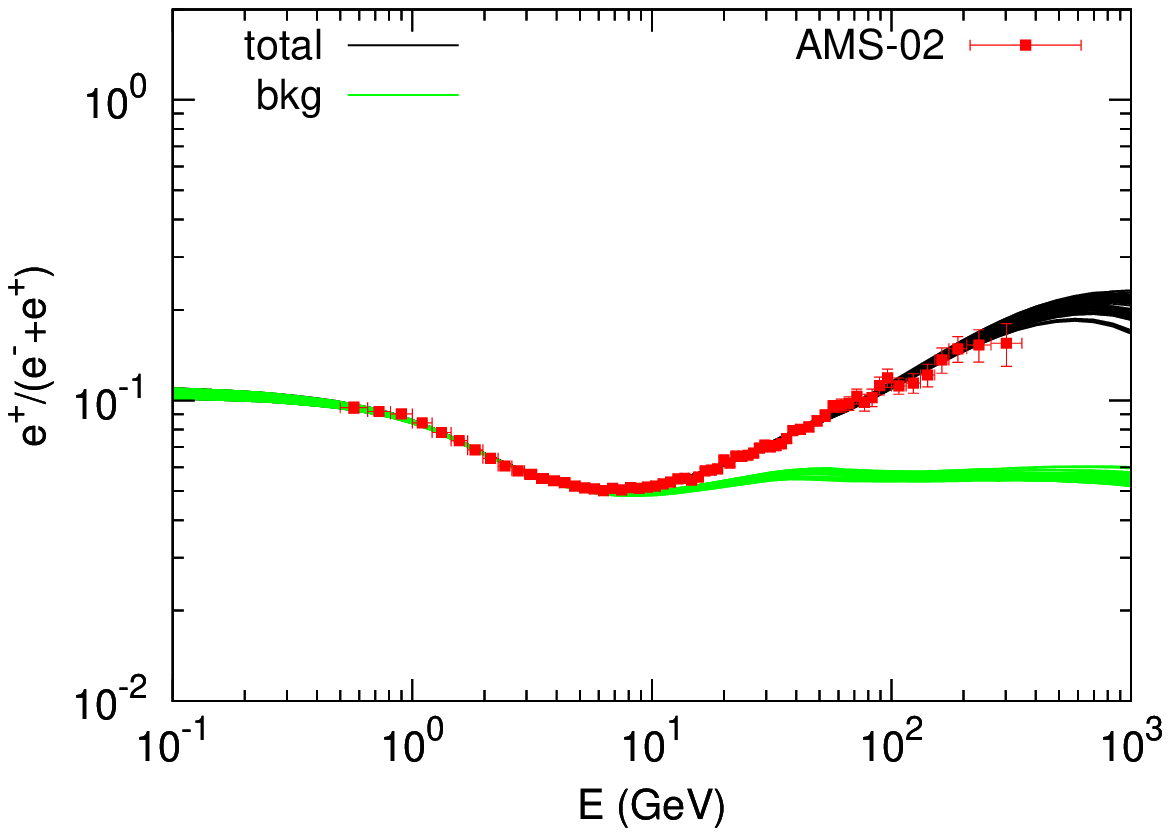}\\
\includegraphics[width=0.5\textwidth]{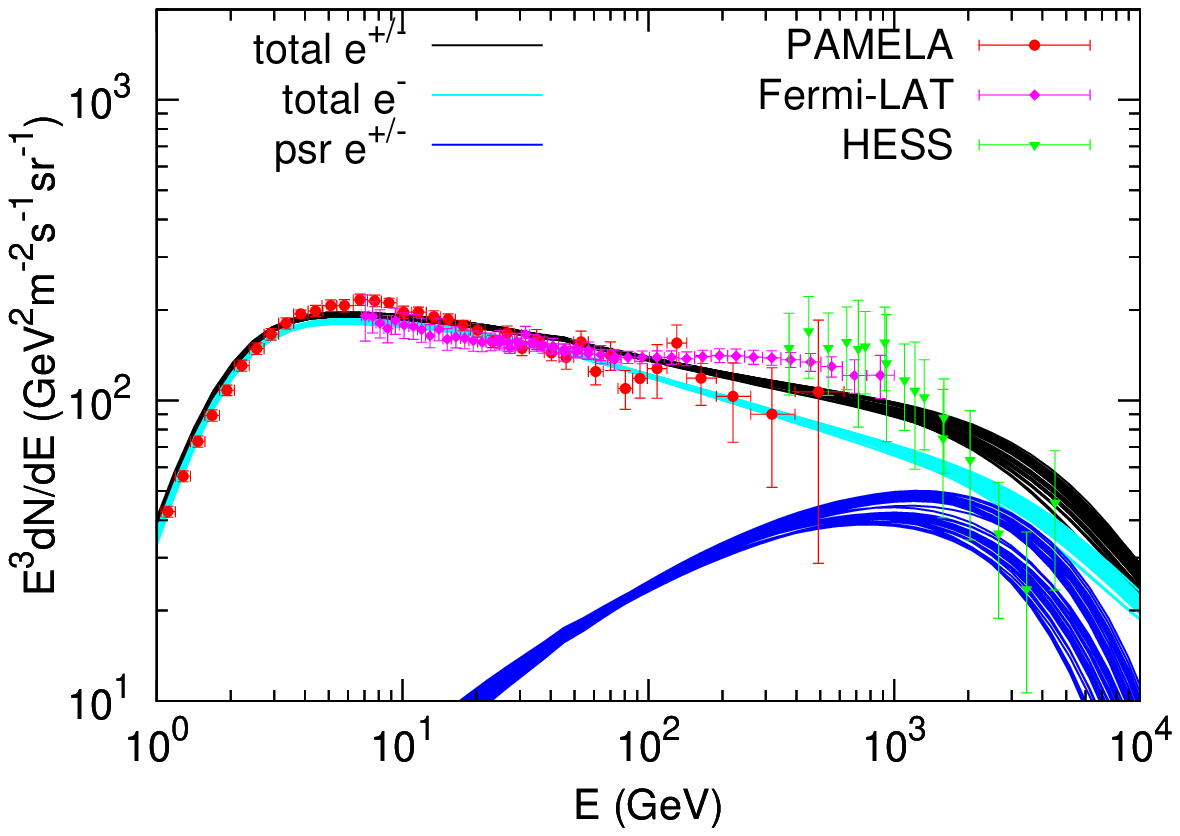}
\includegraphics[width=0.5\textwidth]{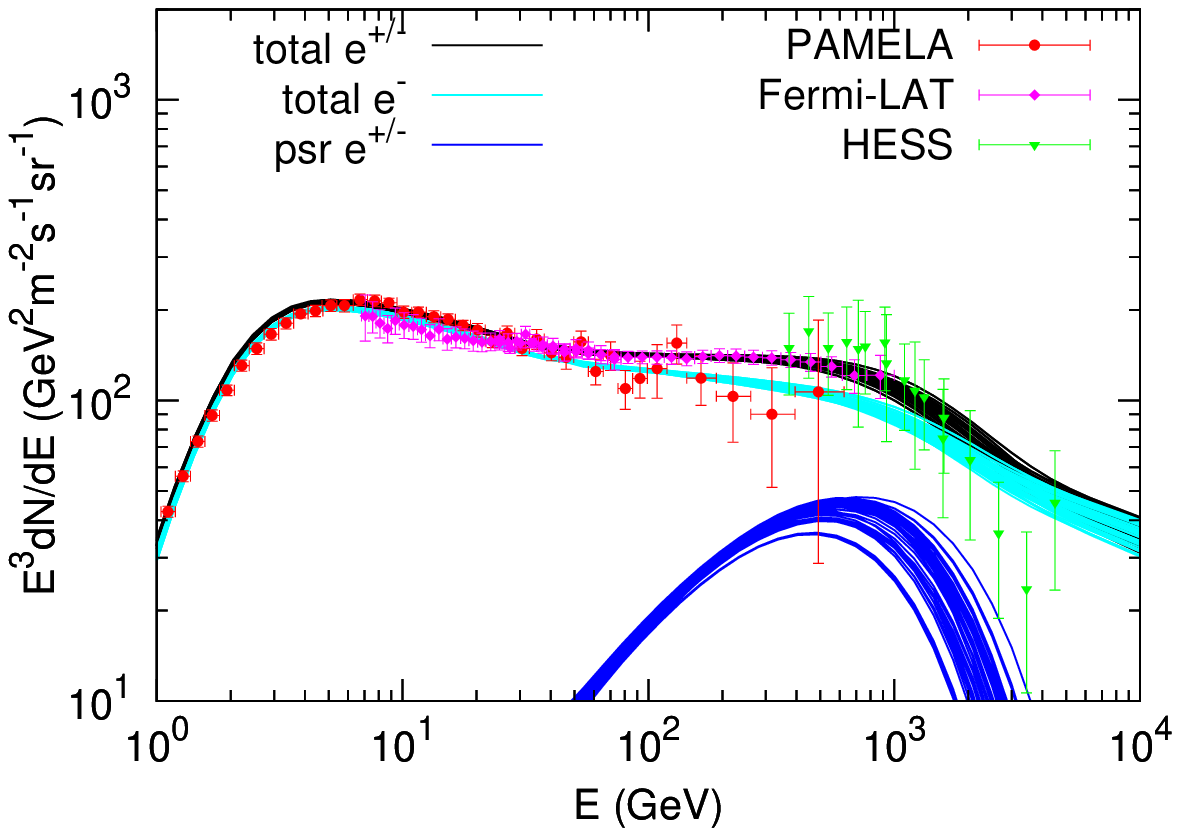}
\caption{The $2\sigma$ ranges of the positron fraction (upper panels)
and electron spectra (lower panels), for 50 randomly selected parameter 
settings of the reference configuration. The left and right panels are
for {\it fittings I} and {\it II}, respectively. The pulsars are adopted 
as the extra sources of the $e^{\pm}$. References of the observational data
are: AMS-02 \cite{2013PhRvL.110n1102A}, PAMELA \cite{2011PhRvL.106t1101A},
Fermi-LAT \cite{2010PhRvD..82i2004A}, and HESS \cite{2008PhRvL.101z1104A,
2009A&A...508..561A}.
\label{fig:ref}}
\end{figure*}

Figure \ref{fig:ref} shows the fitting $2\sigma$ ranges of the positron
fraction (upper panels) and electron (or $e^+e^-$) spectra (lower panels), 
for 50 randomly selected parameter settings of the reference configuration. 
The left panels are for {\it fittings I} (without spectral hardening of the 
primary electrons) and the right panels are for {\it fittings II}. We can 
see that {\it fittings I} seem to under-estimate the high energy $e^+e^-$
fluxes but over-estimate the positron fraction. When a spectral hardening 
of the primary electron flux is introduced, both the positron fraction and 
$e^+e^-$ spectra can be better fitted.

\section{Systematic uncertainties}\label{sys}

In this section we will show the uncertainties when fitting the data
by comparing different model settings with the reference configuration.
The data used in the fittings include the AMS-02 positron fraction
\cite{2013PhRvL.110n1102A}, PAMELA electron spectrum 
\cite{2011PhRvL.106t1101A}, Fermi \cite{2010PhRvD..82i2004A} and HESS
\cite{2008PhRvL.101z1104A,2009A&A...508..561A} total electron/positron 
spectra. The systematic uncertainties of the measurements are added
quadratically to the statistical uncertainties.

\subsection{Propagation parameters}

We first discuss the uncertainties from the propagation parameters. 
It was found that the DR scenario of CR propagation can well describe 
the secondary-to-primary ratios of CR nuclei \cite{2002ApJ...565..280M,
2004ApJ...613..962S}. However, due to the quality of the current CR data, 
the propagation parameters have relatively large uncertainties 
\cite{2011ApJ...729..106T}. In this work we adopt the six groups of 
propagation parameters as reported in Ref. \cite{2012ApJ...761...91A}, 
which are determined through fitting the B/C ratio for six choices of 
the propagation halo height $z_h$ from $2$ to $15$ kpc. These sets of 
propagation parameters embrace most uncertainties from the propagation. 
The values of these parameters are compiled in Table \ref{table:prop}.

\begin{table*}[!htb]
\centering{\small
\caption {Propagation and proton injection parameters}
\begin{tabular}{cccccccccccc}
\hline \hline
 & $D_0^a$ & $z_h$ & $v_A$ & $\delta$ & $dV_c/dz$ & $A_p^b$ & $\gamma_0$ & $\gamma_1$ & $R_{\rm br,1}$ & $\gamma_2$ & $\Phi_p$ \\
 & ($10^{28}$cm$^2$/s) & (kpc) & (km/s) & & (km/s$\cdot$kpc) & & & & (GV) & & (GV) \\
\hline
1 & $2.7$ & $2$ & $35.0$ & $0.33$ & ... & $4.44$ & $1.76$ & $2.43$ & $15.0$ & $2.37$ & $0.32$ \\
2 & $5.3$ & $4$ & $33.5$ & $0.33$ & ... & $4.49$ & $1.79$ & $2.44$ & $13.2$ & $2.37$ & $0.34$ \\
3 & $7.1$ & $6$ & $31.1$ & $0.33$ & ... & $4.51$ & $1.82$ & $2.45$ & $12.9$ & $2.37$ & $0.36$ \\
4 & $8.3$ & $8$ & $29.5$ & $0.33$ & ... & $4.53$ & $1.86$ & $2.46$ & $14.4$ & $2.37$ & $0.36$ \\
5 & $9.4$ & $10$ & $28.6$ & $0.33$ & ... & $4.54$ & $1.87$ & $2.46$ & $14.4$ & $2.38$ & $0.36$ \\
6 & $10.0$ & $15$ & $26.3$ & $0.33$ & ... & $4.51$ & $1.89$ & $2.46$ & $16.3$ & $2.37$ & $0.33$ \\
\hline
7 & $2.5$ & $4$ & ... & $0/0.55^c$ & $6$ & $4.60$ & $2.30$ & $2.37$ & $15.7$ & $2.16$ & $0.42$ \\
\hline
8 & $5.6$ & $6.5$ & $37.6$ & $0.40$ & $11.1$ & $4.50$ & $1.82$ & $2.43$ & $14.5$ & $2.32$ & $0.31$ \\
\hline
\hline
\end{tabular}\vspace{3mm}\\
$^a$Diffusion coefficient at $R=4$ GV.\\
$^b$Normalization at 100 GeV in unit of 
$10^{-9}$cm$^{-2}$s$^{-1}$sr$^{-1}$MeV$^{-1}$.\\
$^c$Below/above $R=4$ GV.
\label{table:prop}
}
\end{table*}

It should be noted that the DR scenario is definitely not the only choice 
to describe the CR propagation. The diffusion plus convection (DC) 
model could also fit the data moderately \cite{2002ApJ...565..280M,
2009PhRvD..79b3512Y}. However, a phenomenological break of the diffusion 
coefficient is needed in order to suppress the low energy B/C ratio 
\cite{2002ApJ...565..280M}. As a comparison, we also take a DC model 
with parameters given in Ref. \cite{2009PhRvD..79b3512Y} in this study.
We note that in Ref. \cite{2010A&A...516A..66P} the diffusion with both
reacceleration and convection (DRC) was shown to be favored, with 
semi-analytical approach of the CR propagation. We have tested such a model 
with the B/C and $^{10}$Be/$^9$Be data. It shows almost no improvement of 
the fitting compared with the DR scenario. The possible reason of the 
difference may come from the difference between the semi-analytical and 
the full numerical approaches of the CR propagation. Nevertheless, 
the expectation of the lepton fluxes in the DRC model might be different 
from the DR or DC models \cite{2009A&A...501..821D}. Therefore we include 
one example of the DRC model in our discussion to illustrate how the 
results may be affected.

\begin{figure*}[!htb]
\includegraphics[width=0.5\textwidth]{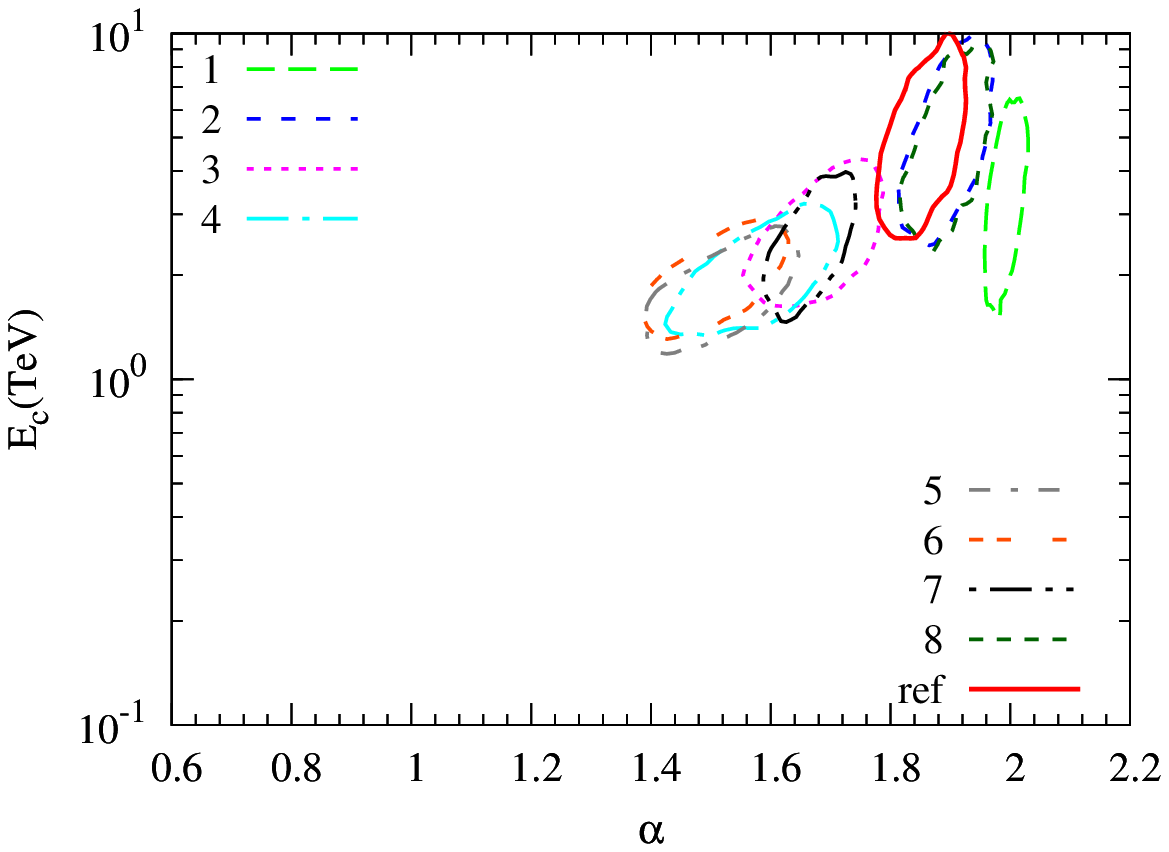}
\includegraphics[width=0.5\textwidth]{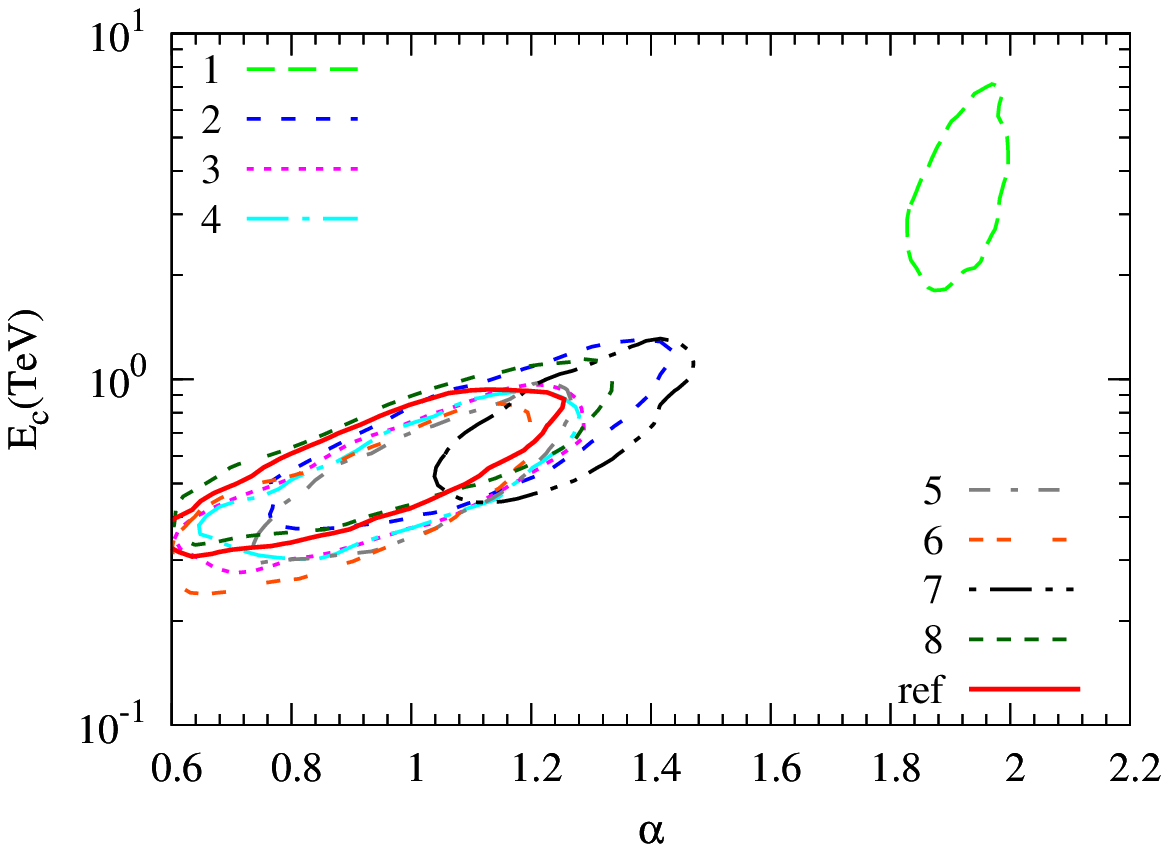}\\
\includegraphics[width=0.5\textwidth]{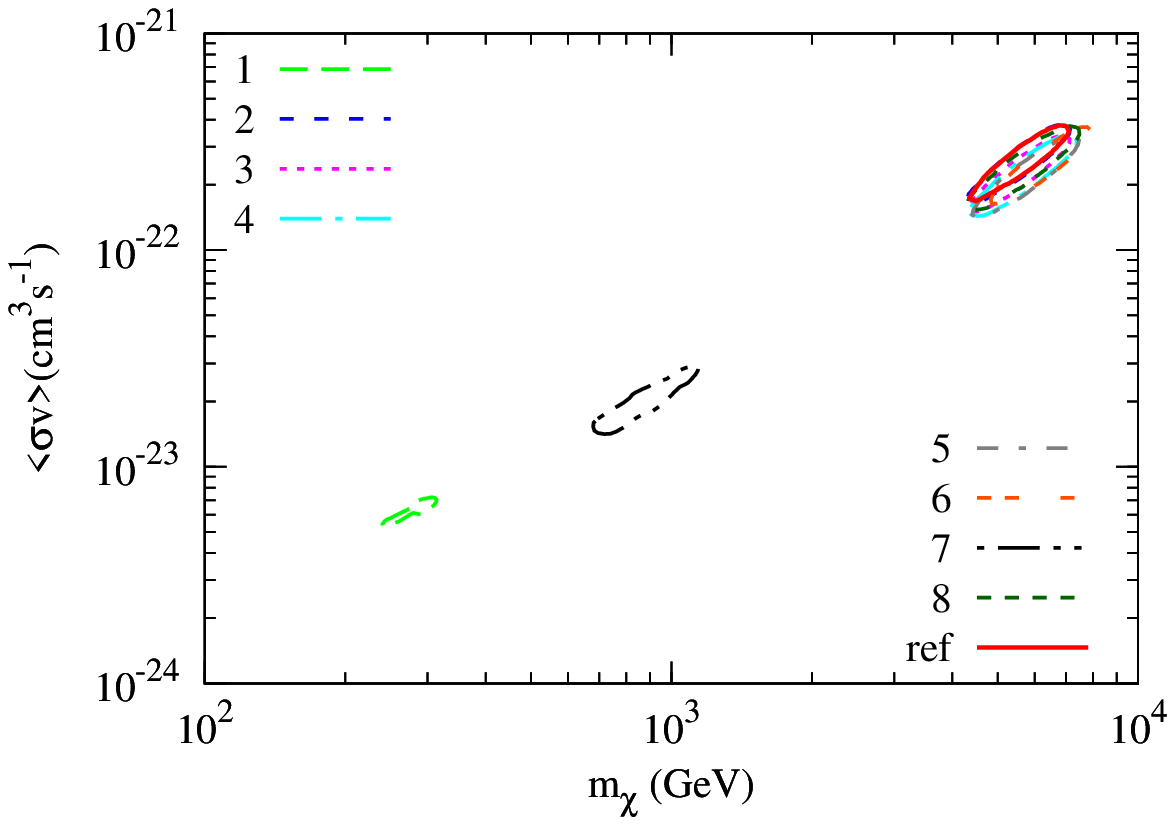}
\includegraphics[width=0.5\textwidth]{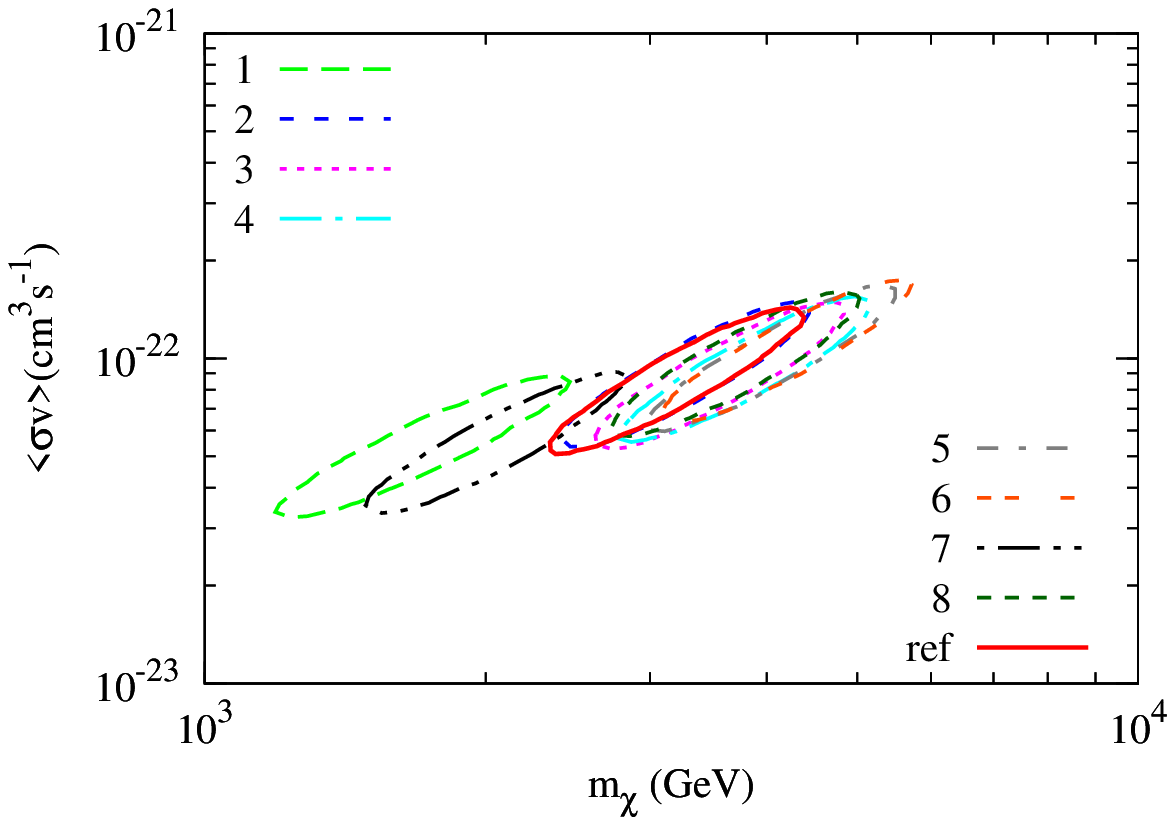}
\caption{Comparison of the $2\sigma$ confidence regions of $\alpha-E_c$
in pulsar scenario (top) and $m_{\chi}-\sv$ in DM scenario (bottom)
for different propagation parameters as given in Table \ref{table:prop}.
The left (right) panels are for the model without (with) spectral hardening 
of the primary electrons, i.e., {\it fittings I} ({\it II}).
\label{fig:prop}}
\end{figure*}

\begin{table*}[!htb]
\centering
\caption {Fitting $\chi^2$ for different propagation parameters.}
\begin{tabular}{cccccccccc}
\hline \hline
 & ref. & 1 & 2 & 3 & 4 & 5 & 6 & 7 & 8\\
\hline
psr-{\it I} (166$^a$) & 398.2 & 476.4 & 344.0 & 342.3 & 359.5 & 378.1 & 414.9 & 313.3 & 369.7 \\
psr-{\it II} (164) & 134.0 & 275.4 & 134.5 & 146.9 & 180.6 & 213.6 & 239.1 & 155.4 & 143.0 \\
\hline
DM-{\it I} (167) & 568.0 & 1261 & 570.9 & 415.9 & 394.4 & 395.8 & 424.8 & 596.4 & 588.7 \\
DM-{\it II} (165) & 133.9 & 425.9 & 135.0 & 149.5 & 187.0 & 221.7 & 252.7 & 163.6 & 145.8 \\
\hline
\hline
\end{tabular}\vspace{2mm}\\
$^a$In parenthesis is the number of dof.
\label{table:chi2_prop}
\end{table*}

The best-fitting $\chi^2$ values for the reference configuration and the 
above mentioned 8 groups of propagation parameters are summarized in 
Table \ref{table:chi2_prop}. 
From these results we see that in general {\it fittings I} can hardly 
fit the data, neither for the pulsar model nor for the DM model. If a
spectral hardening of the primary electrons is included ({\it fittings 
II}), the fitting results improve significantly. Except for the propagation
parameters with extreme values of $z_h$, i.e. parameter settings 1,
5 and 6, the reduced $\chi^2$ values are all smaller or close to $1$
for {\it fittings II}. The results imply strongly that a high energy 
spectral hardening of the primary electrons is needed. The physical 
implication of the electron spectral hardening will be discussed later 
in Sec. IV.

Figure \ref{fig:prop} shows the $2\sigma$ confidence regions on the 
$\alpha-E_c$ plane for the pulsar scenario ($m_{\chi}-\sv$ plane for
the DM scenario). The left panels are for {\it fittings I} and the right
panels are for {\it fittings II}. It is shown that for {\it fittings I} 
the contours diverse very significant from each other, while for 
{\it fittings II} they are more convergent. From Table 
\ref{table:chi2_prop} we can also see that the $\chi^2$ values for 
{\it fittings I} are all too large to be good fittings. It implies that 
the resulting contours derived in {\it fittings I} are less statistically 
meaningful. Nonetheless, the parameters for {\it fittings II} also have 
some dispersion among different propagation parameters. It is most 
remarkable for the parameter setting 1, which actually gives the worst 
fitting to the data among those parameter settings. For those with 
acceptable fitting results ($\chi^2$/dof $\sim1$) the favored parameter 
regions do not differ much. Roughly speaking, the parameter regions 
of the extra sources may enlarge by a factor of $\sim2$ compared with the
reference configuration\footnote{Taking the DM scenario for example,
the reference configuration gives roughly $2-4$ TeV for $m_{\chi}$, and
it spans $\sim1-5$ TeV for different propagation models. The constraint 
on $m_{\chi}$ becomes looser by a factor of 2.}, for reasonable 
choices of the propagation parameters. Finally, we note that the 
results of DC and DRC models can be covered by the wide range of the 6 
DR models.

\subsection{Low energy data selection}

The low energy part (lower than tens of GeV) of the $e^+e^-$ spectrum 
contains very complicated physics because many effects enter in this 
energy region. The reacceleration/convection, solar modulation and 
the injection spectral index will all affect the $e^+e^-$ spectra. At 
the same time, the measurement uncertainties are the smallest in the 
low energy range, which contribute significantly to the fittings. 
As most studies are interested in the properties of the extra sources 
which contribute to the high energy part, the low energy data below 
$\sim10$ GeV are simply excluded in many works to avoid the complexity. 
In this subsection, we discuss how the exclusion of the low energy data 
may affect the conclusions of the study.

\begin{table*}[!htb]
\centering
\caption {Fitting $\chi^2$/dof for different selections of the data.}
\begin{tabular}{cccccc}
\hline \hline
 & ref. & $E>5$ GeV & $E>10$ GeV & $E>20$ GeV & Fermi ($E>70$ GeV)\\
\hline
psr-{\it I}  & 398.2/166 & 163.5/140 & 118.7/122 & 61.0/97 & 220.7/129 \\
psr-{\it II} & 134.0/164 & 110.2/138 & 89.7/120 & 48.7/95 & 75.2/127 \\
\hline
DM-{\it I}  & 568.0/167 & 316.7/141 & 158.2/123 & 67.3/98 & 393.4/130 \\
DM-{\it II} & 133.9/165 & 110.3/139 & 90.2/121 & 48.7/96 & 75.5/128 \\
\hline
\hline
\end{tabular}
\label{table:chi2_low}
\end{table*}

We repeat the fittings of the reference configuration with dropping the 
data below 5, 10 and 20 GeV and check how the fitting results change. 
The resulting $\chi^2$ values are shown in Table \ref{table:chi2_low}. 
We find that the higher the data selection threshold, the smaller the 
differences between {\it fittings I} and {\it II}. For the pulsar scenario 
with $E>5$ GeV, {\it fittings II} are slightly favored, and for $E>10$ 
and $E>20$ GeV cases, both {\it fittings I} and {\it II} give acceptable 
description to the data. Similar conclusion can also be drawn for the DM 
scenario. Such a result is reasonable because the requirement of a
spectral hardening of the primary electrons actually comes from the wide
band constraints from the data. 

In Ref. \cite{2013JCAP...12..011G} since only the data above 20 GeV are 
employed they can fit all the data without introducing an extra break at 
primary electron spectrum. This is consistent with our results. 
The reason should be that the low energy data from PAMELA favors a 
different electron spectrum from that favored by the Fermi-LAT data. 
If the low energy data are excluded, there are less constraints from 
PAMELA, and Fermi-LAT data will dominate the behavior of the background 
electrons. In this case a single power-law ($R^{-\gamma_1}$) will be enough 
to fit the data.

Since the low energy $e^+e^-$ spectra of Fermi-LAT show direct discrepancy 
with that of PAMELA and AMS-02 \cite{2013ICRC-AMS02}, we also test the 
case with only the Fermi-LAT data above $\sim70$ GeV. It is shown that 
dropping the low energy data of Fermi-LAT does not help improve the fitting
if no spectral hardening of the primary electrons is assumed ({\it fittings 
I}). {\it Fittings II} are still highly favored in this case (see Table 
\ref{table:chi2_low}).

\begin{figure*}[!htb]
\includegraphics[width=0.5\textwidth]{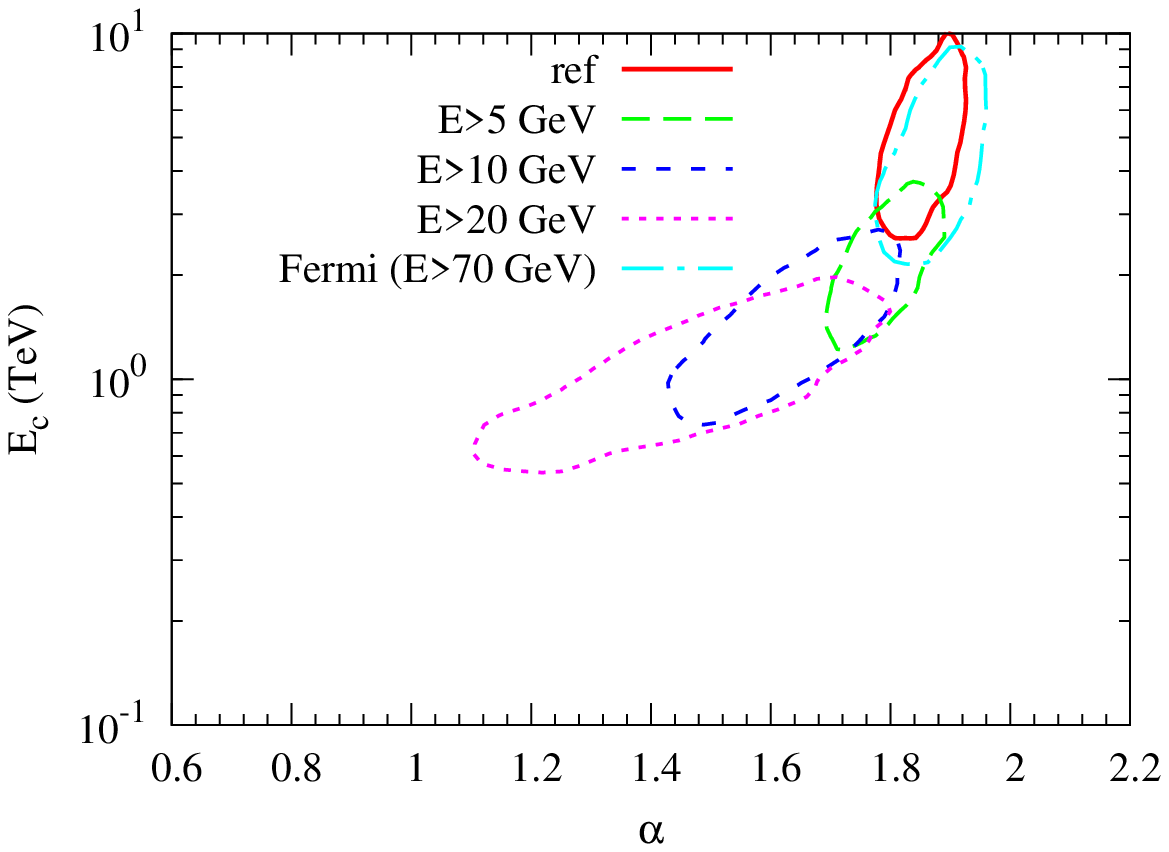}
\includegraphics[width=0.5\textwidth]{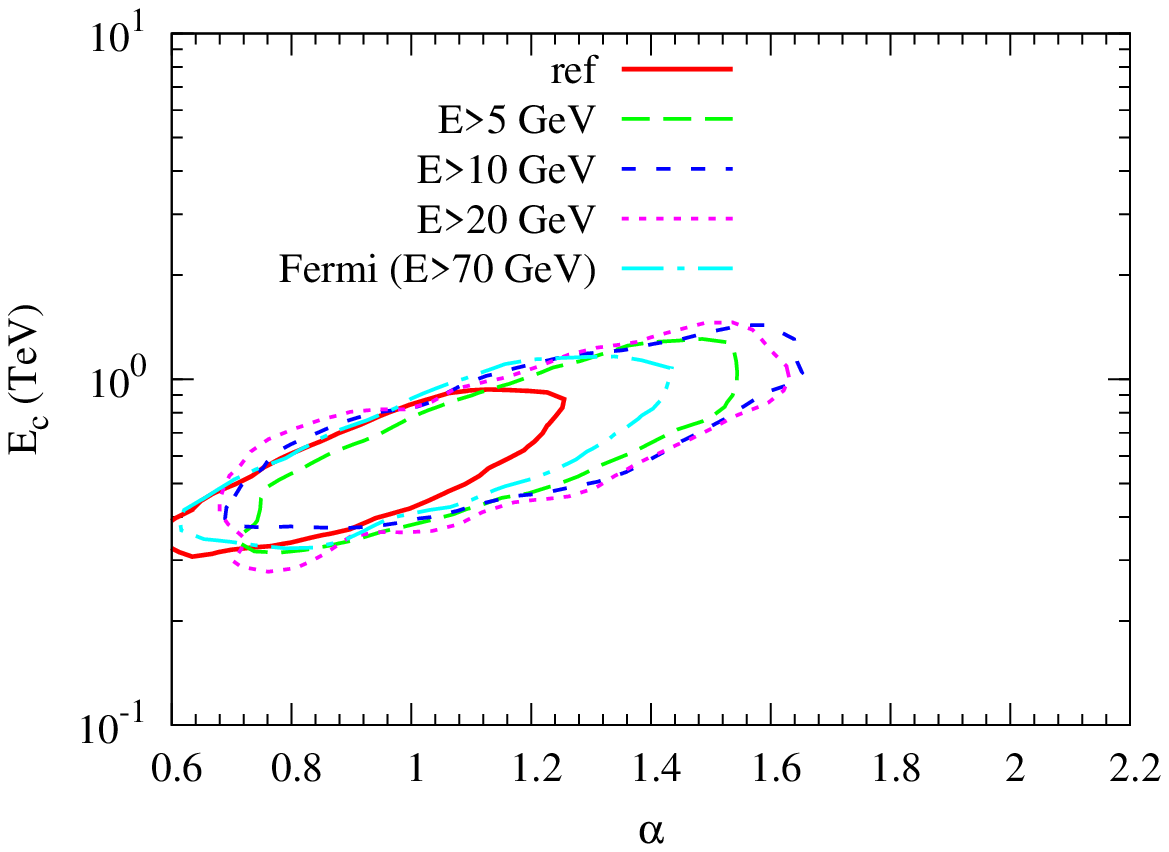}\\
\includegraphics[width=0.5\textwidth]{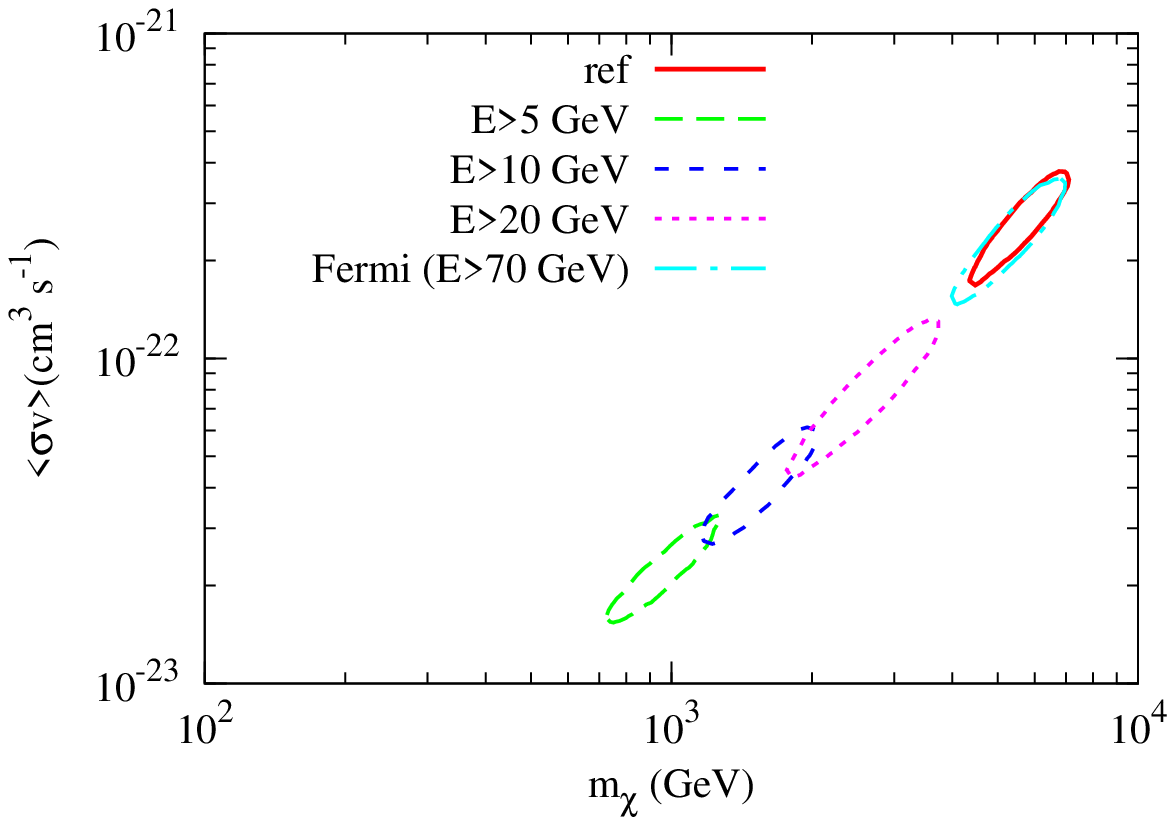}
\includegraphics[width=0.5\textwidth]{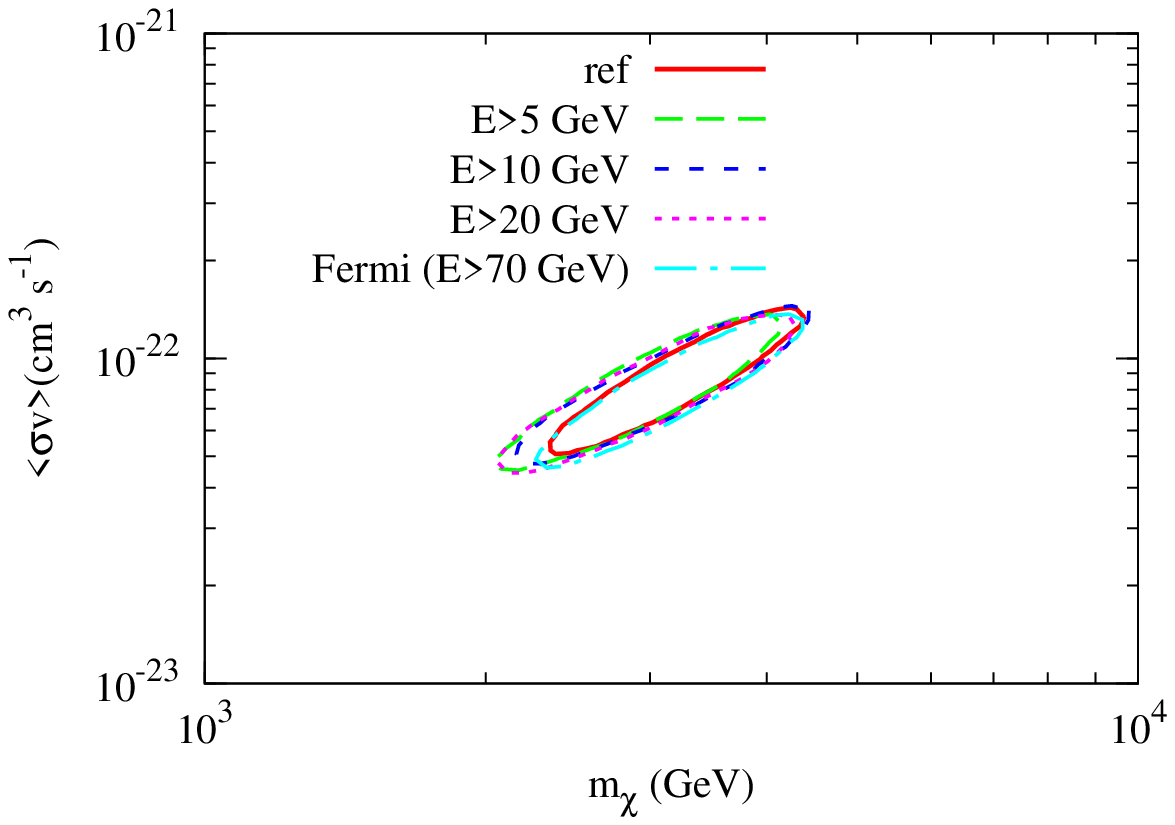}
\caption{Same as Figure \ref{fig:prop} but for comparisons of different
selections of the low energy data as given in Table \ref{table:chi2_low}.
\label{fig:low}}
\end{figure*}

The $2\sigma$ confidence regions of $\alpha-E_c$ and $m_{\chi}-\sv$
for these fittings are given in Figure \ref{fig:low}. It is shown that for 
{\it fittings I} the parameter regions differ significantly from each 
other due to the poor fittings. For {\it fittings II} the favored parameter 
regions converge much better for different cases. 

\subsection{Solar modulation}

As shown in the last subsection the low energy data are important and
may affect the fitting results. However, the low energy CRs are affected 
by the solar modulation, which still has large uncertainties. Recent
works developed three-dimensional, time and charge-sign dependent solar
modulation model, HELIOPROP, which can reproduce the low energy spectra 
of various kinds of species \cite{2013PhRvL.110h1101M}. It has been
shown that for the PAMELA data taking time, the force-field approximation 
gives very similar results to the three-dimensional model, but for the
AMS-02 data taking time the force-field approximation may not work well
\cite{2014PhRvD..89h3007G}. In this subsection we show how the solar 
modulation affects the fitting results. For simplicity, we keep in the
framework of force-field approximation, but with different choices of 
the modulation potentials to partially take into account the effects 
of different data taking time by different detectors and the 
difference between the charge-sign.

Through fitting the proton data the modulation potential was estimated to 
be about 450 to 550 MV during the working period of PAMELA 
\cite{2011Sci...332...69A}. Such a result has been confirmed by the 
measurement of the diffuse $\gamma$-rays in the solar neighborhood 
with Fermi-LAT \cite{2009ApJ...703.1249A,Casandjian2013ICRC}. 
The periods for electron data taking by PAMELA and Fermi-LAT are almost 
the same in the solar cycle, thus they should share a common modulation 
potential as above. In the reference configuration we have taken the solar 
modulation potential to be independent of the value determined from the 
proton data. Here we apply a Gaussian prior of $\phi=500\pm50$ MV on the 
modulation potential and redo the fitting. As can be seen from Table 
\ref{table:chi2_solar}, the fitting $\chi^2$ values become much larger 
than that of the reference configuration. In the reference configuration, 
the best-fitting value of $\phi$ is about 1000 MV, which differs 
significantly from the above value induced from the $\gamma$-ray and CR 
nuclei observations. Such a discrepancy may imply that the simple force 
field approximation may not be enough to describe the solar modulation 
of all types of CR particles.

\begin{table}[!htb]
\centering
\caption{Fitting $\chi^2$/dof for different treatments of the solar 
modulation.}
\begin{tabular}{ccccc}
\hline \hline
 & ref. & $\phi$ prior & $\phi_1/\phi_2$ & $\phi_+/\phi_-$ \\
\hline
psr-{\it I}   & 398.2/166 & 434.8/166 & 315.6/165 & 313.5/165 \\
psr-{\it II}  & 134.0/164 & 227.0/164 & 127.7/163 & 135.8/163 \\
\hline
DM-{\it I}   & 568.0/167 & 670.9/167 & 525.0/166 & 372.2/166 \\
DM-{\it II}  & 133.9/165 & 228.6/165 & 126.7/164 & 134.3/164 \\
\hline
\hline
\end{tabular}
\label{table:chi2_solar}
\end{table}

\begin{table}[!htb]
\centering
\caption{Mean values and $1\sigma$ uncertainties of the modulation 
potentials (in unit of MV) corresponding to different tests in Table 
\ref{table:chi2_solar}.}
\begin{tabular}{ccccc}
\hline \hline
 & ref. & $\phi$ prior & $\phi_1/\phi_2$ & $\phi_+/\phi_-$ \\
\hline
psr-{\it I}   & $824\pm16$ & $792\pm15$ & $691\pm20$, $1483\pm16^a$ & $688\pm18$, $900\pm16$ \\
psr-{\it II}  & $1001\pm13$ & $965\pm14$ & $928\pm42$, $1341\pm196^a$ & $1060\pm103$, $1002\pm14$ \\
\hline
DM-{\it I}   & $1020\pm10$ & $998\pm10$ & $936\pm13$, $1480\pm21^a$ & $725\pm22$, $998\pm11$ \\
DM-{\it II}  & $998\pm12$ & $974\pm11$ & $934\pm35$, $1346\pm177^a$ & $1044\pm110$, $999\pm13$ \\
\hline
\hline
\end{tabular}\vspace{2mm}\\
$^a$Parameter is close to the upper limit $1500$ MV of the scan.
\label{table:phi_solar}
\end{table}

\begin{figure*}[!htb]
\includegraphics[width=0.5\textwidth]{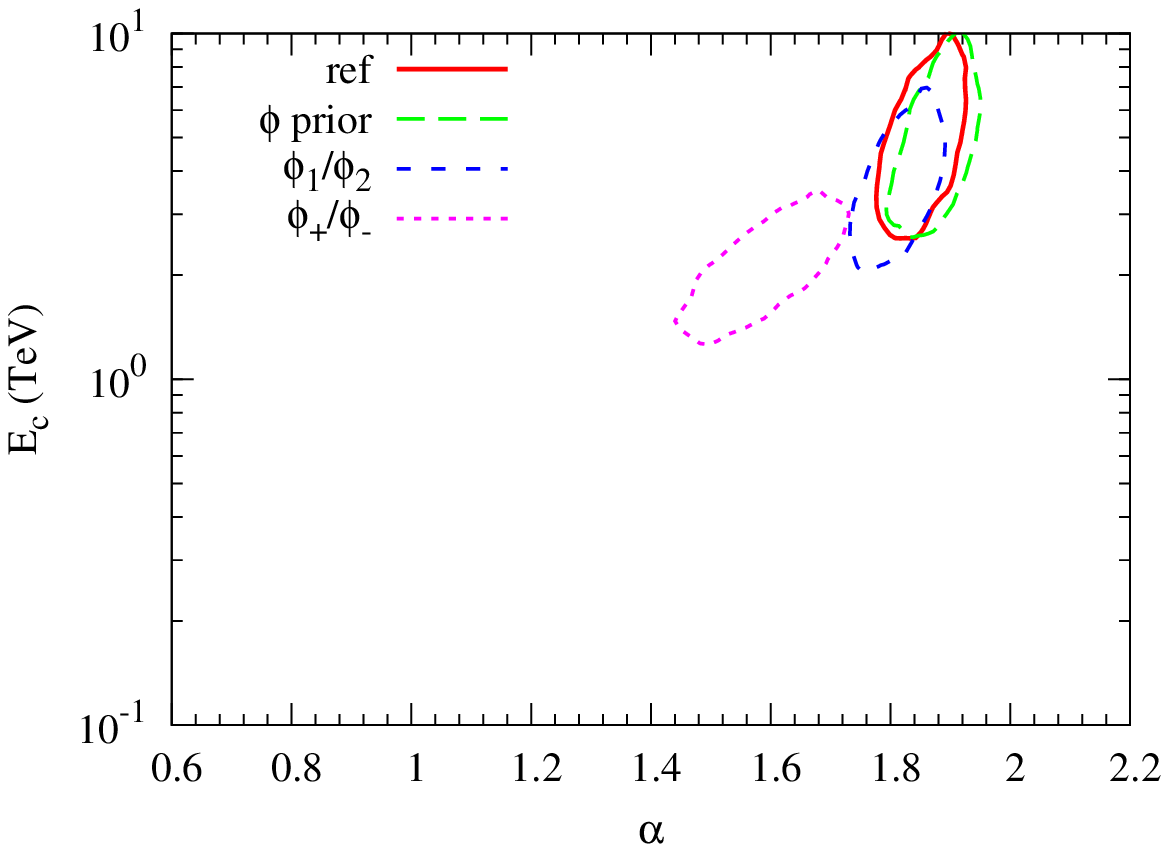}
\includegraphics[width=0.5\textwidth]{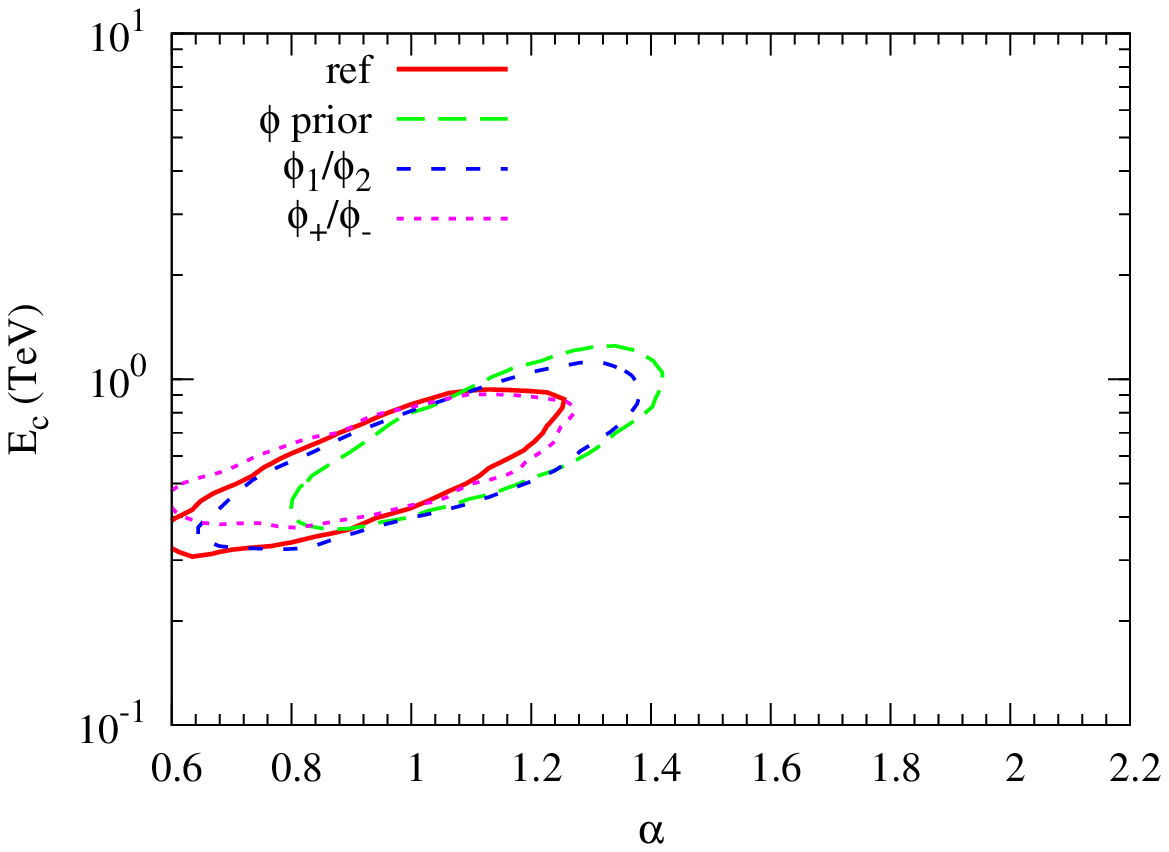}\\
\includegraphics[width=0.5\textwidth]{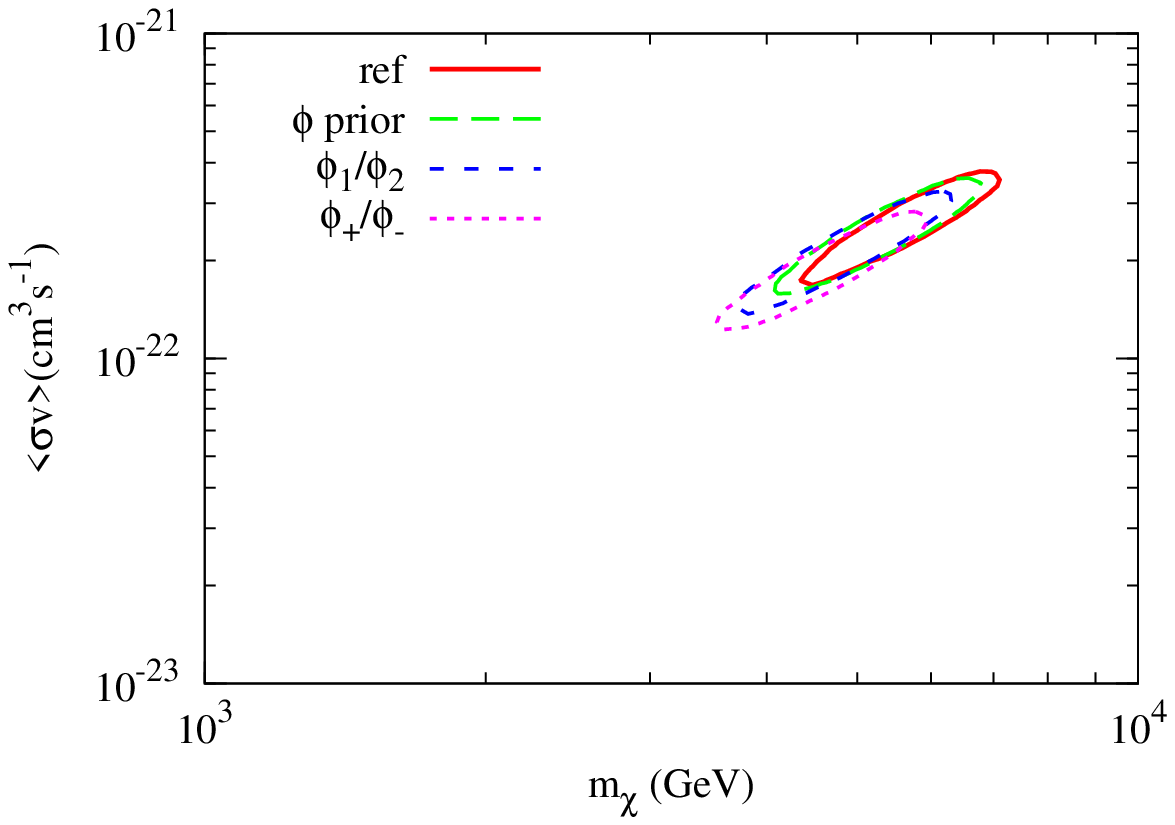}
\includegraphics[width=0.5\textwidth]{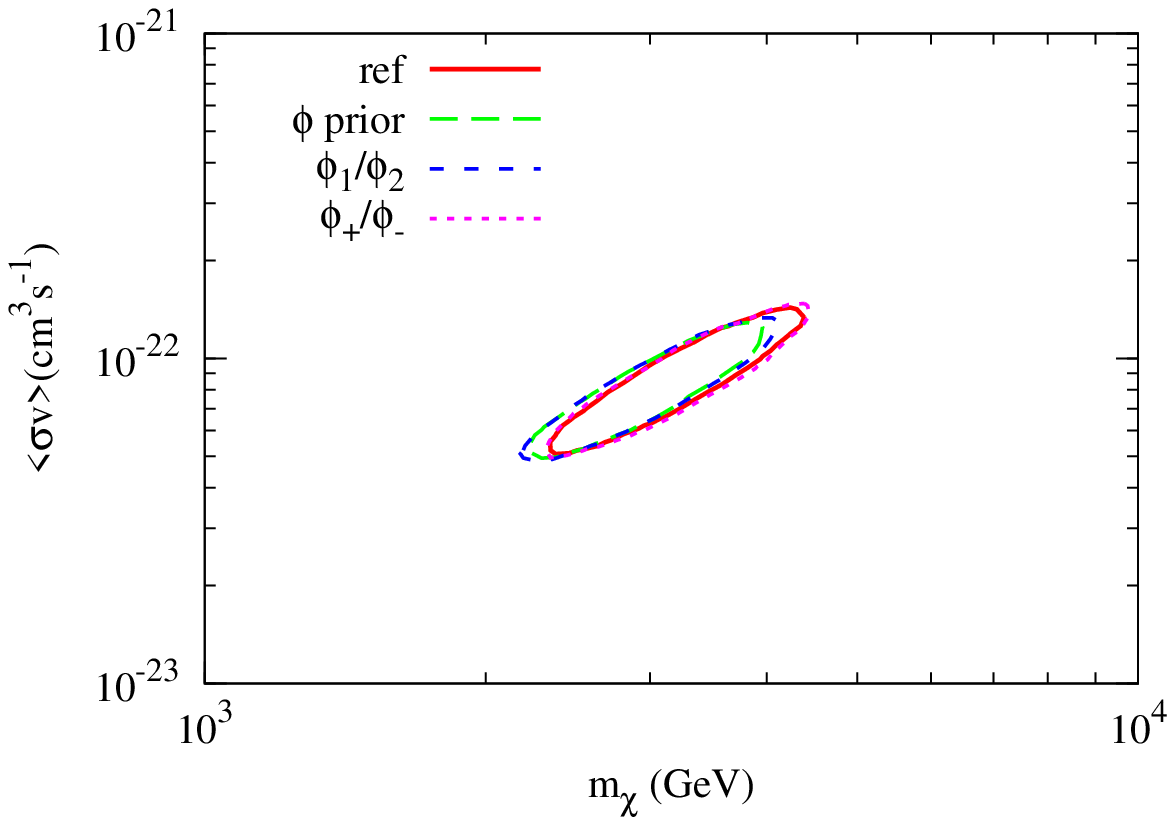}
\caption{Same as Figure \ref{fig:prop} but for comparisons of different
treatments of the solar modulation as given in Table \ref{table:chi2_solar}.
\label{fig:solar}}
\end{figure*}

Furthermore, considering that the data taking period of PAMELA/Fermi-LAT 
is different from that of AMS-02, we adopt two potentials, $\phi_1$ for 
PAMELA/Fermi-LAT and $\phi_2$ for AMS-02 to test whether the fittings can 
be improved. Given one more free parameter, we do have some improvement 
for {\it fittings I} compared with the reference model. However, the 
$\chi^2$ values are still too large (Table \ref{table:chi2_solar}). 
Introducing the spectral hardening of the primary electrons can improve 
the fittings significantly ({\it fittings II}). Compared with the reference 
configuration, the best-fitting $\chi^2$ values become slightly smaller 
for {\it fittings II}. That is to say, a single solar modulation potential 
seems to work well enough under the present model frame, although adding 
another modulation potential will improve the fittings slightly. 
The fitting values of the solar modulation potentials are listed in
Table \ref{table:phi_solar}. We find that in all these fittings, 
$\phi_2>\phi_1$ is shown, which may reflect the fact that AMS-02 works 
close to the solar maximum while PAMELA worked during the moderate phase 
of solar activity.

Finally, to account for the charge-sign dependent modulation effect,
we apply two different modulation potentials, $\phi_+$ and $\phi_-$, on 
positrons and electrons, respectively. Similar with the $\phi_1/\phi_2$
scenario, we have some improvement for {\it fittings I} compared with the 
reference scenario. However, the improvement is not enough to give a good 
fitting. For {\it fittings II}, the best-fitting $\chi^2$ values are almost 
the same as the reference configuration, and there is no improvement with an 
additional free parameter. In addition, we find that $\phi_+\approx\phi_-$
for {\it fittings II} (see Table \ref{table:phi_solar}), which indicates 
that no strong evidence of charge-sign dependent modulation effect is 
needed from the data.

The confidence regions of the extra source parameters for different
treatments of the solar modulation as described above are shown in Figure
\ref{fig:solar}. Similar with the previous cases, we see that for {\it 
fittings I} the parameter regions have relatively large dispersion, and 
for {\it fittings II} they converge much better. 

\subsection{Other uncertainties}

\subsubsection{Hadronic interaction model}

The hadronic interaction model affects the production spectra of secondary 
$e^+e^-$. In the reference configuration we take the hadronic $pp$ collision 
parameterization given in Ref. \cite{2006ApJ...647..692K} (K06). As a test 
we discuss another $pp$ collision model developed in \cite{1986ApJ...307...47D} 
(D86) which combines isobaric model near the threshold 
\cite{1970Ap&SS...6..377S} and scaling representations at high energies 
\cite{1977PhRvD..15..820B}. This model is the default hadronic interaction
model adopted in the GALPROP package (see Ref. \cite{1998ApJ...493..694M} 
for a detailed description).

\begin{figure*}[!htb]
\includegraphics[width=0.5\textwidth]{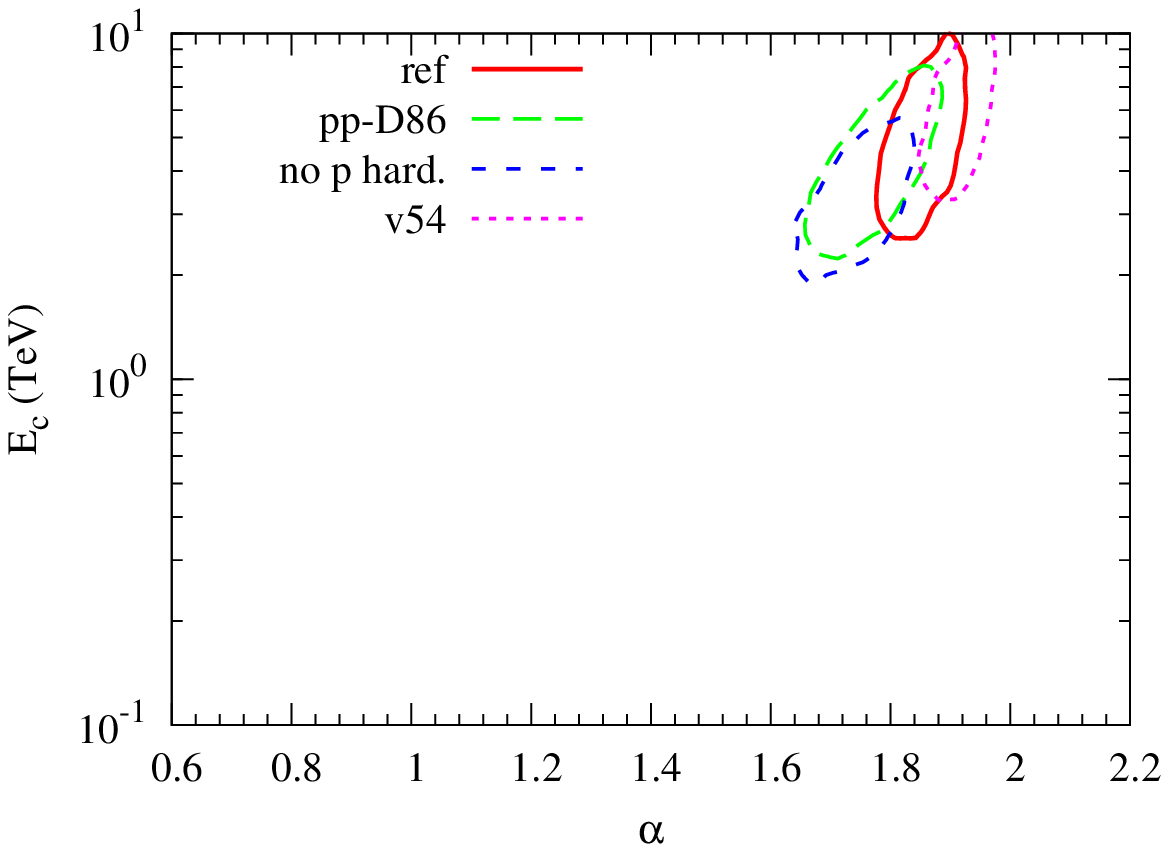}
\includegraphics[width=0.5\textwidth]{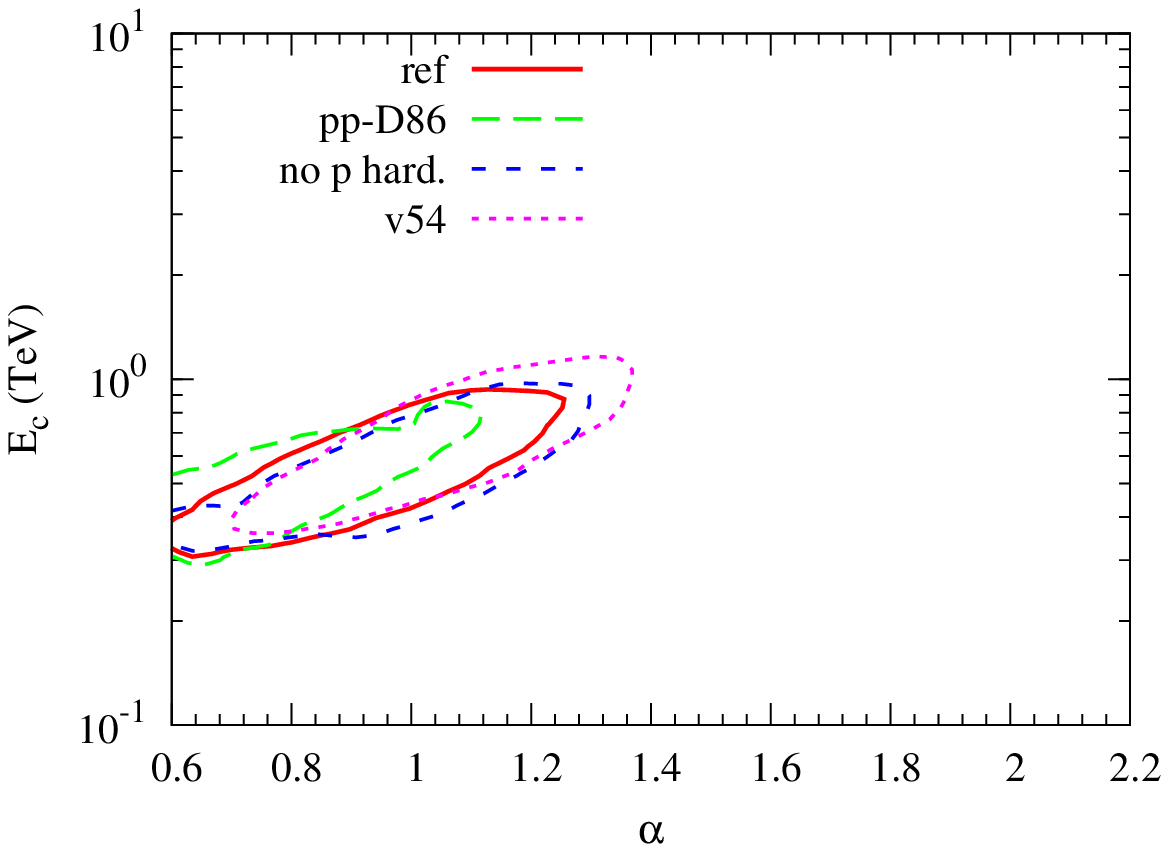}
\includegraphics[width=0.5\textwidth]{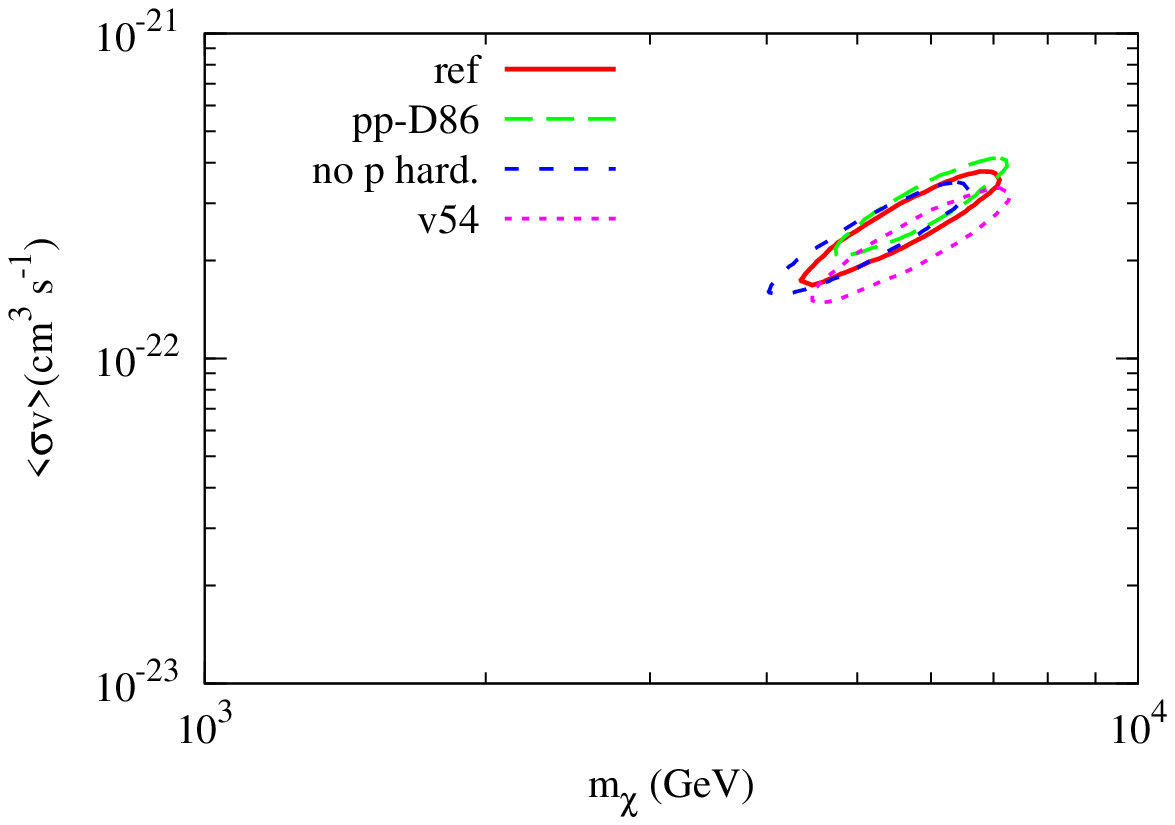}
\includegraphics[width=0.5\textwidth]{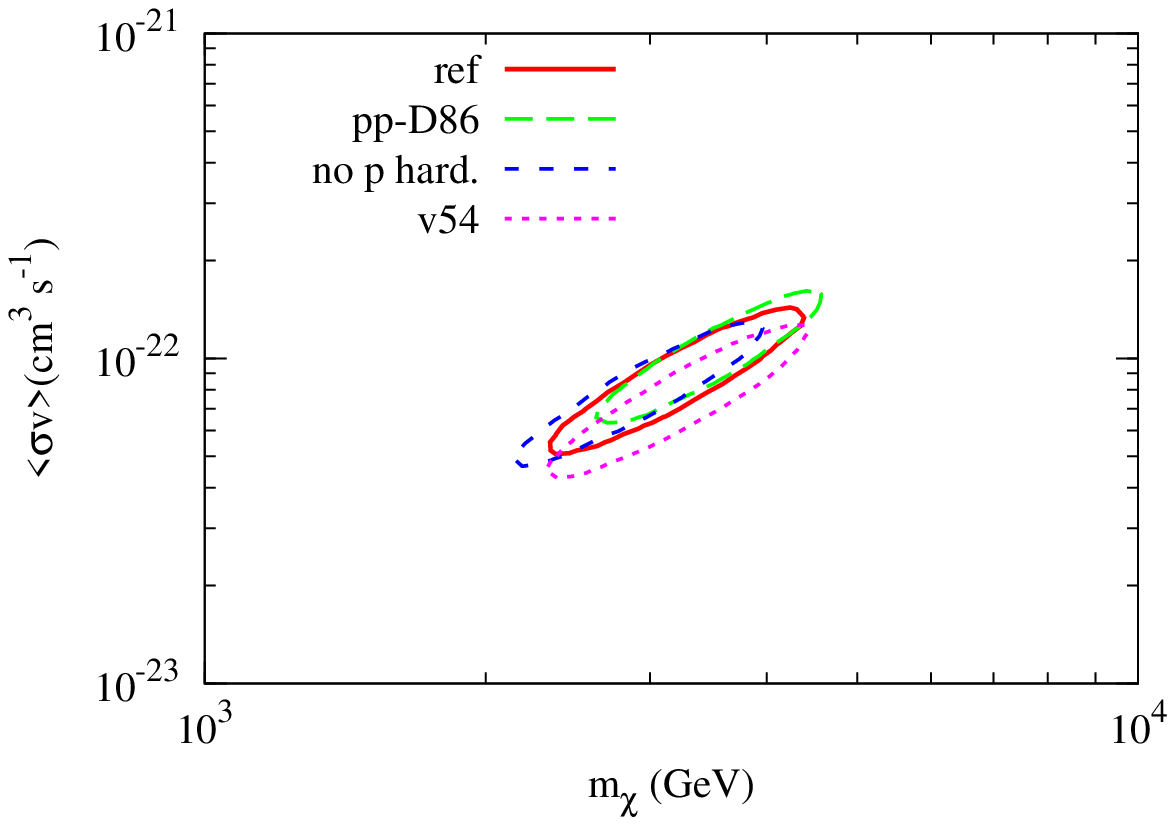}
\caption{Same as Figure \ref{fig:prop} but for comparisons of other
tests as given in Table \ref{table:chi2_other}.
\label{fig:other}}
\end{figure*}

The difference of the expected secondary positron fluxes between the 
two interaction models differs up to several tens of percentages and 
varies with energy \cite{2009A&A...501..821D}. The fitting results are 
presented in Table \ref{table:chi2_other} and Figure \ref{fig:other}. 
As can be seen in Table \ref{table:chi2_other}, the goodness-of-fit of 
the D86 model is generally worse than the K06 model, both for {\it 
fittings I} and {\it II}. However, the favored confidence regions of
the source parameters, as can be seen in Figure \ref{fig:other}, do not 
differ much from each other.

\begin{table}[!htb]
\centering
\caption{Fitting $\chi^2$/dof for other tests.}
\begin{tabular}{ccccc}
\hline \hline
 & ref. & $pp$-D86 & no $p$ hard. & v54 \\
\hline
psr-{\it I}  & 398.2/166 & 505.7/166 & 359.4/166 & 367.9/166 \\
psr-{\it II} & 134.0/164 & 173.5/164 & 137.9/164 & 140.6/164 \\
\hline
DM-{\it I}   & 568.0/167 & 600.3/167 & 486.8/167 & 609.0/167 \\
DM-{\it II}  & 133.9/165 & 175.3/165 & 137.1/165 & 140.7/165 \\
\hline
\hline
\end{tabular}
\label{table:chi2_other}
\end{table}

\subsubsection{Nuclei spectral hardening}

The ATIC2, CREAM and PAMELA measurements show that there is a uniform 
spectral hardening of the CR nuclei spectra at sub-TV rigidities
\cite{2007BRASP..71..494P,2010ApJ...714L..89A,2011Sci...332...69A}.
However, the preliminary data from AMS-02 tend to disfavor the existence 
of the spectral hardening at $\sim200$ GV of the proton and Helium 
spectra as found by PAMELA \cite{2013ICRC-AMS02}. The spectrum of
the primary protons will affect the prediction of the secondary
positron flux \citep{2011MNRAS.414..985L}. Therefore we test the case 
without nuclei spectral hardening here. We fit the propagated 
proton spectrum to the PAMELA/CREAM data. The fitting $\chi^2/{\rm dof}$ 
values are $60.1/85$ ($33.7/84$) for the case without (with) nuclei 
spectral hardening. The differences of the proton spectra, and thus
the secondary positron spectra, between these two models are actually very 
small. We find that the calculated positron flux differ by $\lesssim5\%$ 
at tens of GeV from the reference configuration. It is different from 
Ref. \citep{2011MNRAS.414..985L}, where the model without the spectral 
hardening used for comparison was derived through fitting the low energy 
data only. It is shown from Table \ref{table:chi2_other} that, for 
{\it fittings I} the goodness-of-fitting has a little improvement 
compared with the reference configuration. However, the $\chi^2$ values 
are still too large. For {\it fittings II} the $\chi^2$ values have almost 
no change. The confidence regions of the source parameters are very close 
to that of the reference one, as shown in Figure \ref{fig:other}.

\subsubsection{Versions of GALPROP}

For all the previous studies in this work we adopt ``v50'' of the 
GALPROP code. A new version ``v54'' of the code has been developed
and made public by the authors in recent years, which has remarkable
improvements in several aspects \cite{2009arXiv0907.0559S}. The most 
relevant aspect is the update of the interstellar radiation field, which 
may affect the propagation of leptons. No significant difference of the 
results between ``v50'' and ``v54'' is found\footnote{Here we adopt
version 54.1.984 of the code. The latest updated version can be
obtained at http://sourceforge.net/projects/galprop/}. See Table 
\ref{table:chi2_other} and Figure \ref{fig:other} for the results. 
Our conclusion obtained based on the ``v50'' version code should not be 
affected given the new version of the propagation code.

\section{Discussion about the electron spectral hardening}\label{dis}

From all the above tests, an additional spectral hardening of the primary 
electron spectrum at $O(100)$ GV is strongly favored in order to fit the
$e^+e^-$ data from $\sim$GeV to $\sim$TeV simultaneously. A natural 
explanation of the spectral hardening could be the discreteness 
of the CR sources \cite{2012MNRAS.421.1209T,2013A&A...555A..48B,
2014JCAP...04..006D}. Since the effective propagation length of TeV 
electrons within the cooling time of TeV electrons, which is $O(10^5)$ yrs,
is $\lesssim$ kpc \cite{2013PhRvD..88b3001Y}, the number of CR sources
which can contribute to the local electrons, e.g. supernova remnants, is 
very small. Thus it is likely that one or few nearby sources contribute more 
significant to the high energy part of the electron spectrum and result 
in a harder spectrum than the background extension \cite{2012MNRAS.421.1209T,
2013A&A...555A..48B,2014JCAP...04..006D}.

The fitting under the assumption of continuous CR source distribution
seems indicating the importance of local sources. However, it will be
impossible to incorporate all the discrete sources in the global fitting
procedure, since there will be too many free parameters in that case.
The continuous source assumption could be regarded as the average of
a randomly assigned CR source distribution. The deviation from this
average result may directly indicate the break down of the continuous
source assumption. The studies taking into account the discrete
distribution of sources can be found in Refs. \cite{2013ApJ...772...18L,
2013PhRvD..88b3001Y,2014JCAP...04..006D}.

Alternatively, other mechanisms proposed to explain the CR nuclei
hardening, such as the superposition of multiple components of sources 
\cite{2011PhRvD..84d3002Y,2012APh....35..449E}, the change of injection 
spectrum due to non-linear acceleration process \cite{2013ApJ...763...47P}, 
the propagation effect \cite{2012ApJ...752L..13T,2012PhRvL.109f1101B,
2014A&A...567A..33T} etc. (see \cite{2012ApJ...752...68V} for a 
compilation and implication of these different types of models), may 
also apply for the electrons. However, the hardening of the electrons
seems to be much more significant than that of nuclei, which may
challenge some of the models.

Another possibility is that the contributions to positrons and electrons
from the extra sources are non-equal, i.e., there is a charge asymmetry
of the extra sources. It can be realized in the asymmetric DM scenario 
within the $R$-parity violation supersymmetry framework 
\cite{2013JCAP...10..008F}. 

\section{Conclusion}\label{con}

In this work we give a systematic study on the uncertainties of fitting
the CR $e^+e^-$ data. The potential sources of the uncertainties discussed
include the CR propagation, the selection of low energy data, the solar 
modulation, the hadronic interaction model, the spectral hardening of 
nuclei and the versions of the propagation code. The major conclusions 
are summarized as follows.
\begin{itemize}
\item In general a spectral hardening of the background electron spectrum 
is favored, in spite that there are various kinds of uncertainties from
the CR propagation and the solar modulation. The break energy is about 
$50-100$ GeV, and the change of the spectral index is about $0.3-0.4$. 
The required spectral hardening may indicate the contributions to high 
energy electrons from nearby CR sources. In this case the continuous 
assumption of the source distribution will break down and the fluctuation 
from discrete source(s) dominate the high energy behavior of the primary 
electrons.
\item The propagation models and parameters lead to one of the main 
uncertainties of the fittings. Varying the propagation parameters in a 
wide range allowed by the present CR data will make the constraints on 
the $e^+e^-$ extra source parameters loosen by a factor of $\sim2$. 
\item The exclusion of low energy data will affect the results of 
the fittings. If the low energy data below tens of GeV are excluded in 
the fittings, the AMS-02/Fermi data can be fitted without introducing an 
electron spectral hardening. This is because only the high energy spectral
behavior is constrained by the data. Such a conclusion based on the 
fitting to the high energy data only may be biased.
\item The solar modulation does not affect the fitting results much,
although it does lead to uncertainties in the modeling of the low energy
spectra. The modulation potential for $e^+e^-$ is usually greater than 
that for protons.
\item The hadronic interaction models, the proton spectral hardening and 
the propagation code versions have very small effects in the results.
\item For fittings with large $\chi^2$ values ({\it fittings I} and few 
cases in {\it fittings II}) the derived source parameters have very large 
dispersion, while for the fittings with acceptable $\chi^2$ values the 
parameter regions converge very well for different model settings. 
For the pulsar scenario, the spectral index $\alpha\sim1$ and the cutoff 
energy $E_c\sim0.5$ TeV are found, and for the DM annihilation into a pair 
of tauons, $m_{\chi}\sim3$ TeV and $\sv\sim6\times10^{-23}$ cm$^3$s$^{-1}$ 
are favored. 

\end{itemize}

\acknowledgments
We thank the anonymous referee for helpful comments and suggestions
which improve this paper much.
This work is supported by 973 Program under Grant No. 2013CB837000,
and by National Natural Science Foundation of China under Grant Nos. 
11135009, 11105155 and by the Strategic Priority Research Program
``The Emergence of Cosmological Structures'' of the Chinese Academy 
of Sciences under Grant No. XDB09000000.

\vspace{3mm}
\noindent
{\bf \large Appendix: Results for different annihilation channel of DM}

Here we present the fitting contours on the $m_{\chi}-\sv$ plane for
DM annihilation channels $\mu^+\mu^-$, $\tau^+\tau^-$, $4e$, $4\mu$
and $4\tau$, respectively. For the $4e$, $4\mu$ and $4\tau$ chennels
we assume the mass of the intermediate particle $\phi$ is 100 GeV.
The other settings are the same as the reference configuration plus a 
high energy spectral hardening of the primary electrons as presented 
in Sec. \ref{ref}.

\begin{figure}[!htb]
\centering
\includegraphics[width=0.7\textwidth]{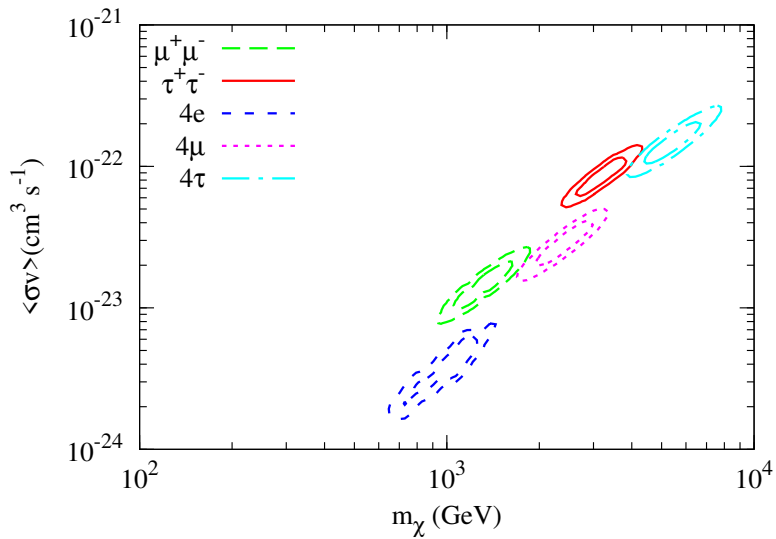}
\caption{$1$ and $2\sigma$ fitting contours on the $m_{\chi}-\sv$ plane
for different DM annihilation channels.
\label{fig:msv_ch}}
\end{figure}

\bibliographystyle{JHEP}
\bibliography{/home/yuanq/work/cygnus/tex/refs}

\providecommand{\href}[2]{#2}\begingroup\raggedright\begin{thebibliography}{10}

\bibitem{1997ApJ...482L.191B}
S.~W. {Barwick}, et~al., {\it {Measurements of the Cosmic-Ray Positron Fraction
  from 1 to 50 GeV}},  {\em \apjl} {\bf 482} (June, 1997) L191,
  [\href{http://xxx.lanl.gov/abs/astro-ph/9703192}{{\tt astro-ph/9703192}}].

\bibitem{2007PhLB..646..145A}
M.~{Aguilar}, et~al., {\it {Cosmic-ray positron fraction measurement from 1 to
  30 GeV with AMS-01}},  {\em Phys. Lett. B} {\bf 646} (Mar., 2007) 145--154,
  [\href{http://xxx.lanl.gov/abs/astro-ph/0703154}{{\tt astro-ph/0703154}}].

\bibitem{2009Natur.458..607A}
O.~{Adriani}, et~al., {\it {An anomalous positron abundance in cosmic rays with
  energies 1.5-100GeV}},  {\em \nat} {\bf 458} (Apr., 2009) 607--609,
  [\href{http://xxx.lanl.gov/abs/0810.4995}{{\tt arXiv:0810.4995}}].

\bibitem{2010APh....34....1A}
O.~{Adriani}, et~al., {\it {A statistical procedure for the identification of
  positrons in the PAMELA experiment}},  {\em Astroparticle Physics} {\bf 34}
  (Aug., 2010) 1--11, [\href{http://xxx.lanl.gov/abs/1001.3522}{{\tt
  arXiv:1001.3522}}].

\bibitem{2013PhRvL.110n1102A}
M.~{Aguilar}, et~al., {\it {First result from the Alpha Magnetic Spectrometer
  on the Internatinal Space Station: precision measurement of the positron
  fraction in primary cosmic rays of 0.5-350 GeV}},  {\em \prl} {\bf 110}
  (Apr., 2013) 141102.

\bibitem{2009PhRvD..79b1302S}
P.~D. {Serpico}, {\it {Possible causes of a rise with energy of the cosmic ray
  positron fraction}},  {\em \prd} {\bf 79} (Jan., 2009) 021302,
  [\href{http://xxx.lanl.gov/abs/0810.4846}{{\tt arXiv:0810.4846}}].

\bibitem{2013NuPhS.243...85M}
I.~V. {Moskalenko}, {\it {Cosmic Rays in the Milky Way and Beyond}},  {\em
  Nuclear Physics B Proceedings Supplements} {\bf 243} (Oct., 2013) 85--91,
  [\href{http://xxx.lanl.gov/abs/1308.5482}{{\tt arXiv:1308.5482}}].

\bibitem{2009MPLA...24.2139H}
X.~{He}, {\it {Dark Matter Annihilation Explanation for $e^{\pm}$ Excesses in
  Cosmic Ray}},  {\em Modern Physics Letters A} {\bf 24} (2009) 2139--2160,
  [\href{http://xxx.lanl.gov/abs/0908.2908}{{\tt arXiv:0908.2908}}].

\bibitem{2010IJMPD..19.2011F}
Y.-Z. {Fan}, B.~{Zhang}, and J.~{Chang}, {\it {Electron/positron Excesses in
  the Cosmic Ray Spectrum and Possible Interpretations}},  {\em International
  Journal of Modern Physics D} {\bf 19} (2010) 2011--2058,
  [\href{http://xxx.lanl.gov/abs/1008.4646}{{\tt arXiv:1008.4646}}].

\bibitem{2012APh....39....2S}
P.~D. {Serpico}, {\it {Astrophysical models for the origin of the positron
  ''excess''}},  {\em Astroparticle Physics} {\bf 39} (Dec., 2012) 2--11,
  [\href{http://xxx.lanl.gov/abs/1108.4827}{{\tt arXiv:1108.4827}}].

\bibitem{2012Prama..79.1021C}
M.~{Cirelli}, {\it {Indirect searches for dark matter}},  {\em Pramana} {\bf
  79} (Nov., 2012) 1021--1043, [\href{http://xxx.lanl.gov/abs/1202.1454}{{\tt
  arXiv:1202.1454}}].

\bibitem{2013FrPhy...8..794B}
X.-J. {Bi}, P.-F. {Yin}, and Q.~{Yuan}, {\it {Status of dark matter
  detection}},  {\em Frontiers of Physics} {\bf 8} (Dec., 2013) 794--827,
  [\href{http://xxx.lanl.gov/abs/1409.4590}{{\tt arXiv:1409.4590}}].

\bibitem{2010PhRvD..81b3516L}
J.~{Liu}, Q.~{Yuan}, X.~J. {Bi}, H.~{Li}, and X.~M. {Zhang}, {\it {Markov chain
  Monte Carlo study on dark matter property related to the cosmic $e^{\pm}$
  excesses}},  {\em \prd} {\bf 81} (Jan., 2010) 023516,
  [\href{http://xxx.lanl.gov/abs/0906.3858}{{\tt arXiv:0906.3858}}].

\bibitem{2012PhRvD..85d3507L}
J.~{Liu}, Q.~{Yuan}, X.-J. {Bi}, H.~{Li}, and X.~{Zhang}, {\it {Cosmic ray
  Monte Carlo: A global fitting method in studying the properties of the new
  sources of cosmic e$^{\pm}$ excesses}},  {\em \prd} {\bf 85} (Feb., 2012)
  043507, [\href{http://xxx.lanl.gov/abs/1106.3882}{{\tt arXiv:1106.3882}}].

\bibitem{2015APh....60....1Y}
Q.~{Yuan}, X.-J. {Bi}, G.-M. {Chen}, Y.-Q. {Guo}, S.-J. {Lin}, and X.~{Zhang},
  {\it {Implications of the AMS-02 positron fraction in cosmic rays}},  {\em
  Astroparticle Physics} {\bf 60} (Jan., 2015) 1--12,
  [\href{http://xxx.lanl.gov/abs/1304.1482}{{\tt arXiv:1304.1482}}].

\bibitem{2011PhRvL.106t1101A}
O.~{Adriani}, et~al., {\it {Cosmic-Ray Electron Flux Measured by the PAMELA
  Experiment between 1 and 625 GeV}},  {\em \prl} {\bf 106} (May, 2011) 201101,
  [\href{http://xxx.lanl.gov/abs/1103.2880}{{\tt arXiv:1103.2880}}].

\bibitem{2009PhRvL.102r1101A}
A.~A. {Abdo}, et~al., {\it {Measurement of the Cosmic Ray $e^{+}+e^{-}$
  Spectrum from 20GeV to 1TeV with the Fermi Large Area Telescope}},  {\em
  \prl} {\bf 102} (May, 2009) 181101,
  [\href{http://xxx.lanl.gov/abs/0905.0025}{{\tt arXiv:0905.0025}}].

\bibitem{2010PhRvD..82i2004A}
M.~{Ackermann}, et~al., {\it {Fermi LAT observations of cosmic-ray electrons
  from 7 GeV to 1 TeV}},  {\em \prd} {\bf 82} (Nov., 2010) 092004,
  [\href{http://xxx.lanl.gov/abs/1008.3999}{{\tt arXiv:1008.3999}}].

\bibitem{2008PhRvL.101z1104A}
F.~{Aharonian}, et~al., {\it {Energy Spectrum of Cosmic-Ray Electrons at TeV
  Energies}},  {\em \prl} {\bf 101} (Dec., 2008) 261104,
  [\href{http://xxx.lanl.gov/abs/0811.3894}{{\tt arXiv:0811.3894}}].

\bibitem{2009A&A...508..561A}
F.~{Aharonian}, et~al., {\it {Probing the ATIC peak in the cosmic-ray electron
  spectrum with H.E.S.S.}},  {\em \aap} {\bf 508} (Dec., 2009) 561--564,
  [\href{http://xxx.lanl.gov/abs/0905.0105}{{\tt arXiv:0905.0105}}].

\bibitem{2013PhRvD..88b3013C}
I.~{Cholis} and D.~{Hooper}, {\it {Dark matter and pulsar origins of the rising
  cosmic ray positron fraction in light of new data from the AMS}},  {\em \prd}
  {\bf 88} (July, 2013) 023013, [\href{http://xxx.lanl.gov/abs/1304.1840}{{\tt
  arXiv:1304.1840}}].

\bibitem{2013PhRvD..87l3003M}
I.~{Masina} and F.~{Sannino}, {\it {Hints of a charge asymmetry in the electron
  and positron cosmic-ray excesses}},  {\em \prd} {\bf 87} (June, 2013) 123003,
  [\href{http://xxx.lanl.gov/abs/1304.2800}{{\tt arXiv:1304.2800}}].

\bibitem{2013JCAP...11..026J}
H.-B. {Jin}, Y.-L. {Wu}, and Y.-F. {Zhou}, {\it {Implications of the first
  AMS-02 measurement for dark matter annihilation and decay}},  {\em \jcap}
  {\bf 11} (Nov., 2013) 26, [\href{http://xxx.lanl.gov/abs/1304.1997}{{\tt
  arXiv:1304.1997}}].

\bibitem{2013ICRC-AMS02}
{AMS-02 collaboration}, {\it {}},  in {\em International Cosmic Ray
  Conference},
  http://www.ams02.org/2013/07/new-results-from-ams-presented-at-icrc-2013/,
  2013.

\bibitem{2010APh....34..274D}
G.~{di Bernardo}, C.~{Evoli}, D.~{Gaggero}, D.~{Grasso}, and L.~{Maccione},
  {\it {Unified interpretation of cosmic ray nuclei and antiproton recent
  measurements}},  {\em Astroparticle Physics} {\bf 34} (Dec., 2010) 274--283,
  [\href{http://xxx.lanl.gov/abs/0909.4548}{{\tt arXiv:0909.4548}}].

\bibitem{2011ApJ...729..106T}
R.~{Trotta}, G.~{J{\'o}hannesson}, I.~V. {Moskalenko}, T.~A. {Porter}, R.~{Ruiz
  de Austri}, and A.~W. {Strong}, {\it {Constraints on Cosmic-ray Propagation
  Models from A Global Bayesian Analysis}},  {\em \apj} {\bf 729} (Mar., 2011)
  106, [\href{http://xxx.lanl.gov/abs/1011.0037}{{\tt arXiv:1011.0037}}].

\bibitem{2014PhLB..728..250F}
L.~{Feng}, R.-Z. {Yang}, H.-N. {He}, T.-K. {Dong}, Y.-Z. {Fan}, and J.~{Chang},
  {\it {AMS-02 positron excess: New bounds on dark matter models and hint for
  primary electron spectrum hardening}},  {\em Physics Letters B} {\bf 728}
  (Jan., 2014) 250--255, [\href{http://xxx.lanl.gov/abs/1303.0530}{{\tt
  arXiv:1303.0530}}].

\bibitem{2013ApJ...772...18L}
T.~{Linden} and S.~{Profumo}, {\it {Probing the Pulsar Origin of the Anomalous
  Positron Fraction with AMS-02 and Atmospheric Cherenkov Telescopes}},  {\em
  \apj} {\bf 772} (July, 2013) 18,
  [\href{http://xxx.lanl.gov/abs/1304.1791}{{\tt arXiv:1304.1791}}].

\bibitem{2013JCAP...12..011G}
D.~{Gaggero} and L.~{Maccione}, {\it {Model independent interpretation of
  recent CR lepton data after AMS-02}},  {\em \jcap} {\bf 12} (Dec., 2013) 11,
  [\href{http://xxx.lanl.gov/abs/1307.0271}{{\tt arXiv:1307.0271}}].

\bibitem{2013JHEP...07..063I}
M.~{Ibe}, S.~{Matsumoto}, S.~{Shirai}, and T.~T. {Yanagida}, {\it {AMS-02
  positrons from decaying Wino in the pure gravity mediation model}},  {\em
  Journal of High Energy Physics} {\bf 7} (July, 2013) 63,
  [\href{http://xxx.lanl.gov/abs/1305.0084}{{\tt arXiv:1305.0084}}].

\bibitem{2009PhRvD..80f3005M}
D.~{Malyshev}, I.~{Cholis}, and J.~{Gelfand}, {\it {Pulsars versus dark matter
  interpretation of ATIC/PAMELA}},  {\em \prd} {\bf 80} (Sept., 2009) 063005,
  [\href{http://xxx.lanl.gov/abs/0903.1310}{{\tt arXiv:0903.1310}}].

\bibitem{2010ApJ...710..958K}
N.~{Kawanaka}, K.~{Ioka}, and M.~M. {Nojiri}, {\it {Is Cosmic Ray Electron
  Excess from Pulsars Spiky or Smooth?: Continuous and Multiple
  Electron/Positron Injections}},  {\em \apj} {\bf 710} (Feb., 2010) 958--963,
  [\href{http://xxx.lanl.gov/abs/0903.3782}{{\tt arXiv:0903.3782}}].

\bibitem{2009PhRvD..79j3513H}
D.~{Hooper}, A.~{Stebbins}, and K.~M. {Zurek}, {\it {Excesses in cosmic ray
  positron and electron spectra from a nearby clump of neutralino dark
  matter}},  {\em \prd} {\bf 79} (May, 2009) 103513,
  [\href{http://xxx.lanl.gov/abs/0812.3202}{{\tt arXiv:0812.3202}}].

\bibitem{2009PhRvD..79l3517K}
M.~{Kuhlen} and D.~{Malyshev}, {\it {ATIC, PAMELA, HESS, and Fermi data and
  nearby dark matter subhalos}},  {\em \prd} {\bf 79} (June, 2009) 123517,
  [\href{http://xxx.lanl.gov/abs/0904.3378}{{\tt arXiv:0904.3378}}].

\bibitem{2008A&A...479..427L}
J.~{Lavalle}, Q.~{Yuan}, D.~{Maurin}, and X.~{Bi}, {\it {Full calculation of
  clumpiness boost factors for antimatter cosmic rays in the light of
  {$\Lambda$}CDM N-body simulation results. Abandoning hope in clumpiness
  enhancement?}},  {\em \aap} {\bf 479} (Feb., 2008) 427--452,
  [\href{http://xxx.lanl.gov/abs/0709.3634}{{\tt arXiv:0709.3634}}].

\bibitem{2009PhRvD..80c5023B}
P.~{Brun}, T.~{Delahaye}, J.~{Diemand}, S.~{Profumo}, and P.~{Salati}, {\it
  {Cosmic ray lepton puzzle in the light of cosmological N-body simulations}},
  {\em \prd} {\bf 80} (Aug., 2009) 035023,
  [\href{http://xxx.lanl.gov/abs/0904.0812}{{\tt arXiv:0904.0812}}].

\bibitem{2013PhLB..727....1Y}
Q.~{Yuan} and X.-J. {Bi}, {\it {Reconcile the AMS-02 positron fraction and
  Fermi-LAT/HESS total e$^{±}$ spectra by the primary electron spectrum
  hardening}},  {\em Physics Letters B} {\bf 727} (Nov., 2013) 1--7,
  [\href{http://xxx.lanl.gov/abs/1304.2687}{{\tt arXiv:1304.2687}}].

\bibitem{1998ApJ...509..212S}
A.~W. {Strong} and I.~V. {Moskalenko}, {\it {Propagation of Cosmic-Ray Nucleons
  in the Galaxy}},  {\em \apj} {\bf 509} (Dec., 1998) 212--228,
  [\href{http://xxx.lanl.gov/abs/astro-ph/9807150}{{\tt astro-ph/9807150}}].

\bibitem{2011A&A...534A..54S}
A.~W. {Strong}, E.~{Orlando}, and T.~R. {Jaffe}, {\it {The interstellar
  cosmic-ray electron spectrum from synchrotron radiation and direct
  measurements}},  {\em \aap} {\bf 534} (Oct., 2011) A54,
  [\href{http://xxx.lanl.gov/abs/1108.4822}{{\tt arXiv:1108.4822}}].

\bibitem{2007BRASP..71..494P}
A.~D. {Panov}, et~al., {\it {Elemental energy spectra of cosmic rays from the
  data of the ATIC-2 experiment}},  {\em Bulletin of the Russian Academy of
  Science, Phys.} {\bf 71} (Apr., 2007) 494--497,
  [\href{http://xxx.lanl.gov/abs/astro-ph/0612377}{{\tt astro-ph/0612377}}].

\bibitem{2010ApJ...714L..89A}
H.~S. {Ahn}, et~al., {\it {Discrepant Hardening Observed in Cosmic-ray
  Elemental Spectra}},  {\em \apjl} {\bf 714} (May, 2010) L89--L93,
  [\href{http://xxx.lanl.gov/abs/1004.1123}{{\tt arXiv:1004.1123}}].

\bibitem{2011Sci...332...69A}
O.~{Adriani}, et~al., {\it {PAMELA Measurements of Cosmic-Ray Proton and Helium
  Spectra}},  {\em Science} {\bf 332} (Apr., 2011) 69--,
  [\href{http://xxx.lanl.gov/abs/1103.4055}{{\tt arXiv:1103.4055}}].

\bibitem{2013JCAP...10..008F}
L.~{Feng} and Z.~{Kang}, {\it {Decaying asymmetric dark matter relaxes the
  AMS-Fermi tension}},  {\em \jcap} {\bf 10} (Oct., 2013) 8,
  [\href{http://xxx.lanl.gov/abs/1304.7492}{{\tt arXiv:1304.7492}}].

\bibitem{2014PhRvD..89e5021G}
C.-Q. {Geng}, D.~{Huang}, and L.-H. {Tsai}, {\it {Imprint of multicomponent
  dark matter on AMS-02}},  {\em \prd} {\bf 89} (Mar., 2014) 055021,
  [\href{http://xxx.lanl.gov/abs/1312.0366}{{\tt arXiv:1312.0366}}].

\bibitem{2013A&A...555A..48B}
G.~{Bernard}, T.~{Delahaye}, Y.-Y. {Keum}, W.~{Liu}, P.~{Salati}, and
  R.~{Taillet}, {\it {TeV cosmic-ray proton and helium spectra in the myriad
  model}},  {\em \aap} {\bf 555} (July, 2013) A48,
  [\href{http://xxx.lanl.gov/abs/1207.4670}{{\tt arXiv:1207.4670}}].

\bibitem{2012MNRAS.421.1209T}
S.~{Thoudam} and J.~R. {H{\"o}randel}, {\it {Nearby supernova remnants and the
  cosmic ray spectral hardening at high energies}},  {\em \mnras} {\bf 421}
  (Apr., 2012) 1209--1214, [\href{http://xxx.lanl.gov/abs/1112.3020}{{\tt
  arXiv:1112.3020}}].

\bibitem{2014JCAP...04..006D}
M.~{Di Mauro}, F.~{Donato}, N.~{Fornengo}, R.~{Lineros}, and A.~{Vittino}, {\it
  {Interpretation of AMS-02 electrons and positrons data}},  {\em \jcap} {\bf
  4} (Apr., 2014) 6, [\href{http://xxx.lanl.gov/abs/1402.0321}{{\tt
  arXiv:1402.0321}}].

\bibitem{2009NuPhB.813....1C}
M.~{Cirelli}, M.~{Kadastik}, M.~{Raidal}, and A.~{Strumia}, {\it
  {Model-independent implications of the $e^{\pm}$, $\bar{p}$ cosmic ray
  spectra on properties of Dark Matter}},  {\em Nuclear Physics B} {\bf 813}
  (May, 2009) 1--21, [\href{http://xxx.lanl.gov/abs/0809.2409}{{\tt
  arXiv:0809.2409}}].

\bibitem{2007APh....27..429H}
C.-Y. {Huang}, S.-E. {Park}, M.~{Pohl}, and C.~D. {Daniels}, {\it {Gamma-rays
  produced in cosmic-ray interactions and the TeV-band spectrum of RX
  J1713.7-3946}},  {\em Astroparticle Physics} {\bf 27} (June, 2007) 429--439,
  [\href{http://xxx.lanl.gov/abs/astro-ph/0611854}{{\tt astro-ph/0611854}}].

\bibitem{2004IAUS..218..105L}
D.~R. {Lorimer}, {\it {The Galactic Population and Birth Rate of Radio
  Pulsars}},  in {\em Young Neutron Stars and Their Environments} ({F.~Camilo
  \& B.~M.~Gaensler}, ed.), vol.~218 of {\em IAU Symposium}, p.~105, 2004.

\bibitem{2013ApJS..208...17A}
A.~A. {Abdo}, et~al., {\it {The Second Fermi Large Area Telescope Catalog of
  Gamma-Ray Pulsars}},  {\em \apjs} {\bf 208} (Oct., 2013) 17,
  [\href{http://xxx.lanl.gov/abs/1305.4385}{{\tt arXiv:1305.4385}}].

\bibitem{2013PhRvD..88b3001Y}
P.-F. {Yin}, Z.-H. {Yu}, Q.~{Yuan}, and X.-J. {Bi}, {\it {Pulsar interpretation
  for the AMS-02 result}},  {\em \prd} {\bf 88} (July, 2013) 023001,
  [\href{http://xxx.lanl.gov/abs/1304.4128}{{\tt arXiv:1304.4128}}].

\bibitem{2010PhRvL.105l1101A}
O.~{Adriani}, et~al., {\it {PAMELA Results on the Cosmic-Ray Antiproton Flux
  from 60 MeV to 180 GeV in Kinetic Energy}},  {\em \prl} {\bf 105} (Sept.,
  2010) 121101, [\href{http://xxx.lanl.gov/abs/1007.0821}{{\tt
  arXiv:1007.0821}}].

\bibitem{2009PhRvL.102g1301D}
F.~{Donato}, D.~{Maurin}, P.~{Brun}, T.~{Delahaye}, and P.~{Salati}, {\it
  {Constraints on WIMP Dark Matter from the High Energy PAMELA {\= p}/p Data}},
   {\em \prl} {\bf 102} (Feb., 2009) 071301,
  [\href{http://xxx.lanl.gov/abs/0810.5292}{{\tt arXiv:0810.5292}}].

\bibitem{2010NuPhB.831..178M}
P.~{Meade}, M.~{Papucci}, A.~{Strumia}, and T.~{Volansky}, {\it {Dark Matter
  interpretations of the $e^{\pm}$ excesses after FERMI}},  {\em Nuclear
  Physics B} {\bf 831} (May, 2010) 178--203,
  [\href{http://xxx.lanl.gov/abs/0905.0480}{{\tt arXiv:0905.0480}}].

\bibitem{1997ApJ...490..493N}
J.~F. {Navarro}, C.~S. {Frenk}, and S.~D.~M. {White}, {\it {A Universal Density
  Profile from Hierarchical Clustering}},  {\em \apj} {\bf 490} (Dec., 1997)
  493, [\href{http://xxx.lanl.gov/abs/astro-ph/9611107}{{\tt
  astro-ph/9611107}}].

\bibitem{1968ApJ...154.1011G}
L.~J. {Gleeson} and W.~I. {Axford}, {\it {Solar Modulation of Galactic Cosmic
  Rays}},  {\em \apj} {\bf 154} (Dec., 1968) 1011.

\bibitem{2012AdSpR..49.1587D}
S.~{Della Torre}, et~al., {\it {Effects of solar modulation on the cosmic ray
  positron fraction}},  {\em Advances in Space Research} {\bf 49} (June, 2012)
  1587--1592.

\bibitem{2013PhRvL.110h1101M}
L.~{Maccione}, {\it {Low Energy Cosmic Ray Positron Fraction Explained by
  Charge-Sign Dependent Solar Modulation}},  {\em \prl} {\bf 110} (Feb., 2013)
  081101, [\href{http://xxx.lanl.gov/abs/1211.6905}{{\tt arXiv:1211.6905}}].

\bibitem{1996ApJ...464..507C}
J.~M. {Clem}, D.~P. {Clements}, J.~{Esposito}, P.~{Evenson}, D.~{Huber},
  J.~{L'Heureux}, P.~{Meyer}, and C.~{Constantin}, {\it {Solar Modulation of
  Cosmic Electrons}},  {\em \apj} {\bf 464} (June, 1996) 507.

\bibitem{2009NJPh...11j5021B}
B.~{Beischer}, P.~{von Doetinchem}, H.~{Gast}, T.~{Kirn}, and S.~{Schael}, {\it
  {Perspectives for indirect dark matter search with AMS-2 using cosmic-ray
  electrons and positrons}},  {\em New Journal of Physics} {\bf 11} (Oct.,
  2009) 105021.

\bibitem{2002ApJ...565..280M}
I.~V. {Moskalenko}, A.~W. {Strong}, J.~F. {Ormes}, and M.~S. {Potgieter}, {\it
  {Secondary Antiprotons and Propagation of Cosmic Rays in the Galaxy and
  Heliosphere}},  {\em \apj} {\bf 565} (Jan., 2002) 280--296,
  [\href{http://xxx.lanl.gov/abs/astro-ph/0106567}{{\tt astro-ph/0106567}}].

\bibitem{2004ApJ...613..962S}
A.~W. {Strong}, I.~V. {Moskalenko}, and O.~{Reimer}, {\it {Diffuse Galactic
  Continuum Gamma Rays: A Model Compatible with EGRET Data and Cosmic-Ray
  Measurements}},  {\em \apj} {\bf 613} (Oct., 2004) 962--976,
  [\href{http://xxx.lanl.gov/abs/astro-ph/0406254}{{\tt astro-ph/0406254}}].

\bibitem{2012ApJ...761...91A}
M.~{Ackermann}, et~al., {\it {Constraints on the Galactic Halo Dark Matter from
  Fermi-LAT Diffuse Measurements}},  {\em \apj} {\bf 761} (Dec., 2012) 91,
  [\href{http://xxx.lanl.gov/abs/1205.6474}{{\tt arXiv:1205.6474}}].

\bibitem{2009PhRvD..79b3512Y}
P.~F. {Yin}, Q.~{Yuan}, J.~{Liu}, J.~{Zhang}, X.~J. {Bi}, S.~H. {Zhu}, and
  X.~M. {Zhang}, {\it {PAMELA data and leptonically decaying dark matter}},
  {\em \prd} {\bf 79} (Jan., 2009) 023512,
  [\href{http://xxx.lanl.gov/abs/0811.0176}{{\tt arXiv:0811.0176}}].

\bibitem{2010A&A...516A..66P}
A.~{Putze}, L.~{Derome}, and D.~{Maurin}, {\it {A Markov Chain Monte Carlo
  technique to sample transport and source parameters of Galactic cosmic rays.
  II. Results for the diffusion model combining B/C and radioactive nuclei}},
  {\em \aap} {\bf 516} (June, 2010) A66,
  [\href{http://xxx.lanl.gov/abs/1001.0551}{{\tt arXiv:1001.0551}}].

\bibitem{2009A&A...501..821D}
T.~{Delahaye}, R.~{Lineros}, F.~{Donato}, N.~{Fornengo}, J.~{Lavalle},
  P.~{Salati}, and R.~{Taillet}, {\it {Galactic secondary positron flux at the
  Earth}},  {\em \aap} {\bf 501} (July, 2009) 821--833,
  [\href{http://xxx.lanl.gov/abs/0809.5268}{{\tt arXiv:0809.5268}}].

\bibitem{2014PhRvD..89h3007G}
D.~{Gaggero}, L.~{Maccione}, D.~{Grasso}, G.~{Di Bernardo}, and C.~{Evoli},
  {\it {PAMELA and AMS-02 e$^{+}$ and e$^{-}$ spectra are reproduced by
  three-dimensional cosmic-ray modeling}},  {\em \prd} {\bf 89} (Apr., 2014)
  083007, [\href{http://xxx.lanl.gov/abs/1311.5575}{{\tt arXiv:1311.5575}}].

\bibitem{2009ApJ...703.1249A}
A.~A. {Abdo}, et~al., {\it {Fermi LAT Observation of Diffuse Gamma Rays
  Produced Through Interactions Between Local Interstellar Matter and
  High-energy Cosmic Rays}},  {\em \apj} {\bf 703} (Oct., 2009) 1249--1256,
  [\href{http://xxx.lanl.gov/abs/0908.1171}{{\tt arXiv:0908.1171}}].

\bibitem{Casandjian2013ICRC}
J.-M. {Casandjian}, {\it {}},  in {\em International Cosmic Ray Conference},
  www.cbpf.br/$\sim$icrc2013/papers/icrc2013-0966.pdf, 2013.

\bibitem{2006ApJ...647..692K}
T.~{Kamae}, N.~{Karlsson}, T.~{Mizuno}, T.~{Abe}, and T.~{Koi}, {\it
  {Parameterization of {$\gamma$}, $e^{+/-}$, and Neutrino Spectra Produced by
  p-p Interaction in Astronomical Environments}},  {\em \apj} {\bf 647} (Aug.,
  2006) 692--708, [\href{http://xxx.lanl.gov/abs/astro-ph/0605581}{{\tt
  astro-ph/0605581}}].

\bibitem{1986ApJ...307...47D}
C.~D. {Dermer}, {\it {Binary collision rates of relativistic thermal plasmas.
  II - Spectra}},  {\em \apj} {\bf 307} (Aug., 1986) 47--59.

\bibitem{1970Ap&SS...6..377S}
F.~W. {Stecker}, {\it {The Cosmic {$\gamma$}-Ray Spectrum from Secondary
  Particle Production in Cosmic-Ray Interactions}},  {\em \apss} {\bf 6} (Mar.,
  1970) 377--389.

\bibitem{1977PhRvD..15..820B}
G.~D. {Badhwar}, R.~L. {Golden}, and S.~A. {Stephens}, {\it {Analytic
  representation of the proton-proton and proton-nucleus cross-sections and its
  application to the sea-level spectrum and charge ratio of muons}},  {\em
  \prd} {\bf 15} (Feb., 1977) 820--831.

\bibitem{1998ApJ...493..694M}
I.~V. {Moskalenko} and A.~W. {Strong}, {\it {Production and Propagation of
  Cosmic-Ray Positrons and Electrons}},  {\em \apj} {\bf 493} (Jan., 1998) 694,
  [\href{http://xxx.lanl.gov/abs/astro-ph/9710124}{{\tt astro-ph/9710124}}].

\bibitem{2011MNRAS.414..985L}
J.~{Lavalle}, {\it {Impact of the spectral hardening of TeV cosmic rays on the
  prediction of the secondary positron flux}},  {\em \mnras} {\bf 414} (June,
  2011) 985--991, [\href{http://xxx.lanl.gov/abs/1011.3063}{{\tt
  arXiv:1011.3063}}].

\bibitem{2009arXiv0907.0559S}
A.~W. {Strong}, I.~V. {Moskalenko}, T.~A. {Porter}, G.~{J{\'o}hannesson},
  E.~{Orlando}, and S.~W. {Digel}, {\it {The GALPROP Cosmic-Ray Propagation
  Code}},  {\em ArXiv e-prints} (July, 2009)
  [\href{http://xxx.lanl.gov/abs/0907.0559}{{\tt arXiv:0907.0559}}].

\bibitem{2011PhRvD..84d3002Y}
Q.~{Yuan}, B.~{Zhang}, and X.-J. {Bi}, {\it {Cosmic ray spectral hardening due
  to dispersion in the source injection spectra}},  {\em \prd} {\bf 84} (Aug.,
  2011) 043002, [\href{http://xxx.lanl.gov/abs/1104.3357}{{\tt
  arXiv:1104.3357}}].

\bibitem{2012APh....35..449E}
A.~D. {Erlykin} and A.~W. {Wolfendale}, {\it {A New Component of Cosmic
  Rays?}},  {\em Astroparticle Physics} {\bf 35} (Jan., 2012) 449--456,
  [\href{http://xxx.lanl.gov/abs/1111.3191}{{\tt arXiv:1111.3191}}].

\bibitem{2013ApJ...763...47P}
V.~{Ptuskin}, V.~{Zirakashvili}, and E.-S. {Seo}, {\it {Spectra of Cosmic-Ray
  Protons and Helium Produced in Supernova Remnants}},  {\em \apj} {\bf 763}
  (Jan., 2013) 47, [\href{http://xxx.lanl.gov/abs/1212.0381}{{\tt
  arXiv:1212.0381}}].

\bibitem{2012ApJ...752L..13T}
N.~{Tomassetti}, {\it {Origin of the Cosmic-Ray Spectral Hardening}},  {\em
  \apjl} {\bf 752} (June, 2012) L13,
  [\href{http://xxx.lanl.gov/abs/1204.4492}{{\tt arXiv:1204.4492}}].

\bibitem{2012PhRvL.109f1101B}
P.~{Blasi}, E.~{Amato}, and P.~D. {Serpico}, {\it {Spectral Breaks as a
  Signature of Cosmic Ray Induced Turbulence in the Galaxy}},  {\em Physical
  Review Letters} {\bf 109} (Aug., 2012) 061101,
  [\href{http://xxx.lanl.gov/abs/1207.3706}{{\tt arXiv:1207.3706}}].

\bibitem{2014A&A...567A..33T}
S.~{Thoudam} and J.~R. {H{\"o}randel}, {\it {GeV-TeV cosmic-ray spectral
  anomaly as due to reacceleration by weak shocks in the Galaxy}},  {\em \aap}
  {\bf 567} (July, 2014) A33, [\href{http://xxx.lanl.gov/abs/1404.3630}{{\tt
  arXiv:1404.3630}}].

\bibitem{2012ApJ...752...68V}
A.~E. {Vladimirov}, G.~{J{\'o}hannesson}, I.~V. {Moskalenko}, and T.~A.
  {Porter}, {\it {Testing the Origin of High-energy Cosmic Rays}},  {\em \apj}
  {\bf 752} (June, 2012) 68, [\href{http://xxx.lanl.gov/abs/1108.1023}{{\tt
  arXiv:1108.1023}}].

\end{thebibliography}\endgroup

\end{document}